\documentclass[showpacs,preprintnumbers,amsmath,amssymb]{revtex4}
\usepackage{graphicx}

\newcommand{ \be }{\begin{equation}}
\newcommand{ \ee }{\end{equation}}
\newcommand{ \bea }{\begin{eqnarray}}
\newcommand{ \eea }{\end{eqnarray}}

\usepackage{color}

\begin{document}

\title{Transverse Radial Flow Effects on Two- and Three-Particle Angular Correlations }
\begin{flushright}  \small \sl version 09,  \today \\  \end{flushright} 

\author{Claude A. Pruneau, Sean Gavin , and Sergei A. Voloshin}

\affiliation{Physics and Astronomy Department, Wayne State University, 
Detroit, MI 48152 USA}

\begin{abstract}
We use a simple a transverse radial boost scenario coupled to PYTHIA
events to illustrate the impact radial flow may have of two- and 
three-particle correlation functions measured in heavy-ion collisions. 
We show that modest radial velocities can impart strong modifications 
to the correlation functions, 
some of which may be interpreted as same side ridge and 
away side structure that can mimic conical emission.
\end{abstract}

\maketitle


\section{Introduction}

\begin{figure}[htb]      
\mbox{
\begin{minipage}{0.5\linewidth} \begin{center}
\includegraphics[height=2.4in,width=3.5in]{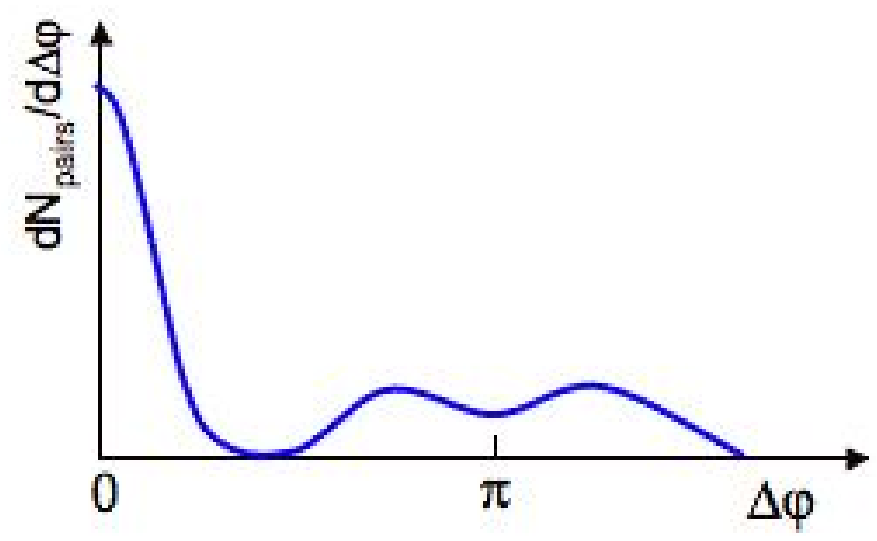}
\end{center} \end{minipage}
\begin{minipage}{0.5\linewidth}\begin{center}
\includegraphics[height=2.4in,width=3.5in]{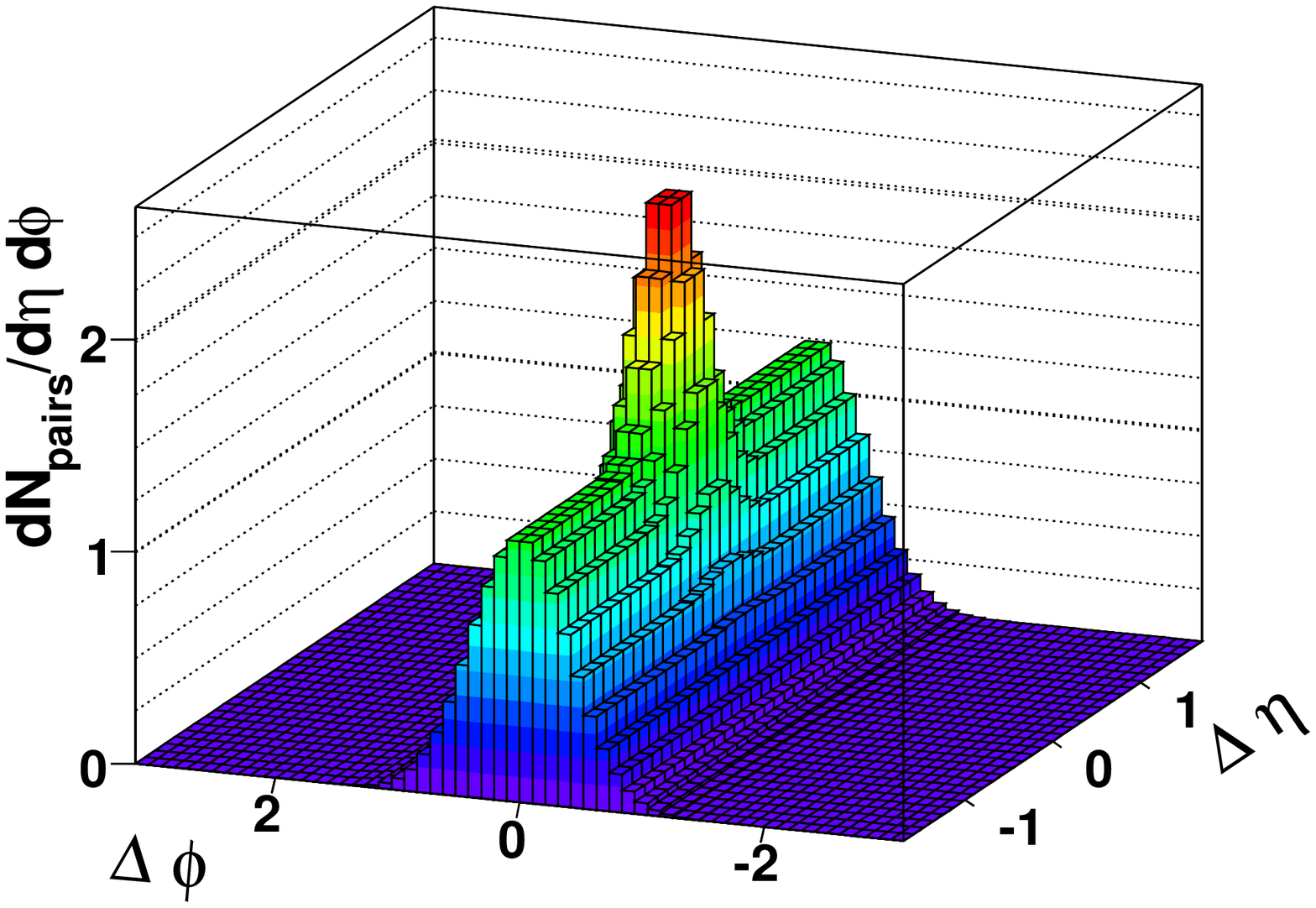}
\end{center}\end{minipage}
}
\caption[]{(Color online)
Cartoon illustrations showing (left) the away-side dip structure observed after elliptic flow subtraction in $\Delta\phi$ correlations, and (right) the ridge structure observed on the near-side in $\Delta\eta$ vs. $\Delta\phi$ correlations.}
\label{fig:Cartoons}
\end{figure}

Observations of $Au+Au$ collisions at the relativistic heavy ion
collider (RHIC), and comparisons with 
results from p+p and d+Au interactions have led to the notion 
that strongly interacting quark gluon 
plasma is formed in $Au+Au$ collisions. 
This conclusion is in part based on the observation of strong 
elliptic flow~\cite{Ackermann01}, dramatic reduction of the high 
transverse momentum ($p_t$) particle production (relative to
expectations from $p+p$ interactions), and suppression of the high transverse 
momentum two-particle back-to-back correlations~\cite{Adler03}. 
Observations of $Au + Au$ collisions have also lead to discoveries of
new, rather unexpected features, which await satisfactory
interpretation.  Among these, we note the observation of a flattening, 
or formation of a dip on the away-side in two particle azimuthal 
angle correlations~\cite{Horner06}, and the discovery of a strong 
ridge-like structure on the near-side of two particle correlations 
measured in relative pseudo-rapidity vs azimuthal angle space~\cite{Putschke06}.
The away side dip and ridge are illustrated qualitatively as cartoons in Figure \ref{fig:Cartoons}.

The observation of a dip at $180^o$ in flow subtracted 
two-particle azimuthal correlations measured in Au + Au collisions 
was reported by both STAR \cite{Star05,Ulery05}, and PHENIX \cite{Phenix06}. 
St\"ocker  argued in~\cite{Stoecker05} the propagation of jets
through matter, and associated energy losses may generate shock 
fronts and deflected jets. 
Shock fronts and wake fields calculations by authors  
of~\cite{Solana05,Ruppert05,Stocker07} suggest the observed 
away-side structures might indeed result from a Mach cone effect. 
Heinz {\em et al.} however have argued, on the basis of 2-D hydro 
calculations, the jet energy loss process is unlikely to result 
in massive away-side structures \cite{Heinz06}. 
Other works suggest the dip-like structure 
might result from large angle gluon
radiation~\cite{Vitev05,Salgado05}, 
Cerenkov gluon radiation~\cite{Dremin05,Majumber05}, or medium drag 
effects~\cite{Salgado05a}. Note also studies based on the AMPT model 
which suggest that parton transport may only partly 
explain the observed two-particle correlations \cite{Ma06}. 
There are thus many suggested explanations for the observed away-side 
structures found in two-particle correlations. 
It is conceivable that discrimination of these different production 
mechanisms may not possible with two-particle correlations, it might 
however be achieved with three particle correlations.  
Three particle correlation measurement techniques and related issues
have been  discussed by one the authors~\cite{Pruneau06} and 
others~\cite{Borghini06, Ajitanand06,Ulery06}. 
Three-particle correlation studies aimed at resolving the 
above production mechanisms are works in 
progress~\cite{Pruneau06a,Ajitanand06,CeresQM06}.

The near side ridge-like structure was first reported by 
the STAR collaboration \cite{Magestro04}, and has been subject 
of extensive studies \cite{Putschke06}.  
The ridge yield for a given associate particle $p_{t}$ range is essentially
independent of the trigger $p_{t}$. 
Additionally, particles in the ridge exhibit $p_{t}$ spectra  
slopes that are independent of the trigger particle momentum. 
This suggests that while the ridge accompanies jet-like particles, 
it may not be strictly caused by jets, but "associated" with
jets. This point of view is perhaps reinforced by the observation 
that the near side $z_T$ di-hadron fragmentation function, 
subtracted for ridge contribution, is independent of the 
trigger momentum~\cite{Putschke06}.  

The near-side ridge has also been subject to numerous theoretical 
investigations. Interpretations of the ridge include induced radiation 
coupled to longitudinal flow \cite{Armesto04}, turbulent color 
fields~\cite{MajumderMuller06}, anisotropic
plasma~\cite{Romatschke07}, a combination of jet quenching and 
strong radial flow~\cite{Voloshin05}, and recombination of locally 
thermal enhanced partons produced by jet energy loss \cite{Hwa05}.
A recent study based on the AMPT event generator \cite{Ma06} indicate significant 
string melting though it produces a ridge like structure cannot
account for the longitudinal extent of the ridge reported by STAR. 

Additionally, it is important to realize that at relatively large
$p_t$, the away-side jet is found to reappear. 
Indeed, STAR reported already at QM05 that away side high $p_t$ particles accompany trigger particles with $p_t>8$ GeV/c~\cite{Magestro06}. 
The away side yield measured in central $Au+Au$ is relatively weaker 
than that found in $d+Au$, but the azimuthal profile of the high $p_t$ away-
side is otherwise unmodified relative to $d+Au$ collisions \cite{Magestro06a}. 

It is thus clear that the structure of jets as measured through 
two-particle correlations is significantly modified in central $Au+Au$
collisions. 
The cause of the modifications and the emergence of new features such 
as the away-side dip and the near-side ridge are however as of yet
unclear. 
While many authors have argued the newly found structures arise from 
jet energy loss and coupling of the jets with gluons of the medium, 
we will here argue that these may be associated with radial flow 
effects. Evidence for elliptical flow is now firmly established, 
and it is widely interpreted as resulting from low viscosity 
hydrodynamics~\cite{EllipticalFlow}. Radial flow must also be present 
given the strong elliptical flow. In fact, particle spectra, as well as HBT correlations are well 
described by Blast-wave parameterizations which include large 
transverse flow velocities (see for
example~\cite{MolnarBarannikova,RetiereLisa} and references therein).
Effects of radial flow on particle correlations have already been
discussed by Voloshin~\cite{Voloshin06} and Pruneau~\cite{Pruneau06}. 

In this work, we study the effects radial flow can impart on two- and three-particle correlations associated with jets, and their underlying events. We neglect string melting, jet quenching, and diffusion processes, and use in-vacuum p+p events, simulated with the PYTHIA event generator, to study the effect of radial flow on produced jets, and the 2-/3-particle correlations they induce. We boost particles produced by PYTHIA radially in the transverse plane, at fixed velocities, and in an arbitrary azimuthal direction. We argue that a form of radial flow drives the formation of the near-side ridge and away-side dip in nuclear collisions. The elliptic and radial flow of particles at soft scales is experimentally well established and well described by hydrodynamics \cite{WhitePapers, Huovinen:2006jp}. The elliptic flow measured in lepton production indicates that charm quarks also flow; see \cite{Zhang:2005ni,adare2007} and references therein. It is therefore not unreasonable to explore how flow might affect the jet production. 

To understand how jets can acquire a flow-like radial motion, consider a quark antiquark pair produced in a hard scattering in a nuclear collision. The center of mass of the pair can have a net transverse momentum at the time it is produced due to initial state parton scattering. In essence, the partons that will ultimately produce the $q\overline{q}$ pair can scatter before the jets are produced.. Such scattering takes place in the production of lepton pairs by the Drell Yan process \cite{Gavin:1988tw},  and is well understood on the basis of pQCD \cite{Guo:1999wy}. This mechanism explains the observed increase of $\langle p_t\rangle$ for lepton pairs produced in pA collisions \cite{Alde:1991sw}, and also contributes to the Cronin effect in jets \cite{Accardi:2005fu}. 
We would expect this contribution to be most pronounced in the $p_t$ region where the Cronin effect is important. If initial state scattering is the primary source of ``jet flow'', then we could expect to see appropriate angular correlations in hard lepton pairs in pA collisions.

We view initial state parton scattering as the least speculative mechanism that may contribute to jet flow. Additional radial motion can be acquired before the hard-scattering if radially-directed color fields are established very early in the collision, as considered by Fries, Kapusta and Li \cite{Fries:2006pv}. 
 This contribution need not be limited to the Cronin region.  The fluid flow of the medium can also add momentum to quark and antiquark pair if the hard scattering that produces the $q\overline{q}$ occurs sufficiently far from the center of the collision volume \cite{RenkRupper:2005}.  For example, flow would push both the $q$ and $\overline q$ radially if they are produced back-to-back in a direction perpendicular to the radius. Jet quenching would accentuate this effect in the correlation function by suppressing the contribution of pairs  in other configurations (i.e., with either the $q$ or $\overline q$ moving radially inward. If this were the dominant mechanism, then we expect the jet flow effect to be correlated with the flow plane.  

The mechanism we discuss here should apply to quark jets of flavors, as well as gluon jets. Electron measurements suggest that charm quarks indeed exhibit elliptic flow; 
see \cite{Zhang:2005ni, adare2007} and references therein. This suggest that flow could also contribute to jet modification.

\begin{figure}[htb]      
\includegraphics[height=3.in,width=4.in]{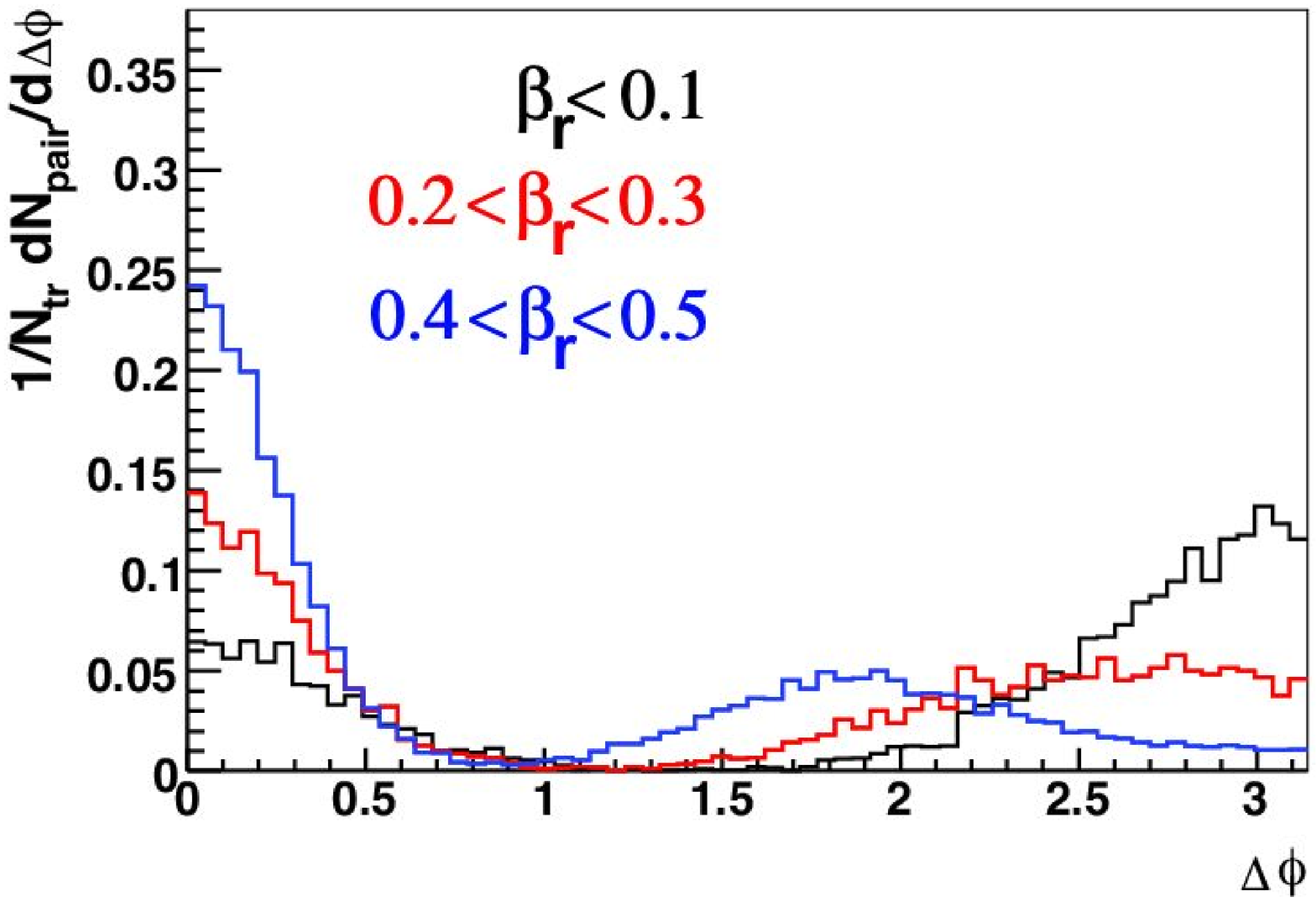}
\caption[]{(Color online) Two-particle azimuthal correlations obtained with a di-jet toy model 
assuming Gaussian azimuthal profile and flat radial boost profiles - described in the text.}
\label{fig:ToyModel}
\end{figure}

As a starting point in this work, and to emphasize the potential role of flow on two particle azimuthal correlations, we show in Figure \ref{fig:ToyModel} results from a di-jet toy model in which the near and away side jets are simulated with Gaussian profiles of 0.4 and 0.5 radians widths respectively. Jet particles are simulated as pions with transverse momentum slope of 300 MeV/c, and have relative probabilities  of 0.7 and 0.3 on the trigger and away sides respectively. Figure \ref{fig:ToyModel} shows correlation functions obtained with flat radial boost profiles for $\beta_r<0.1$ in black, $0.2<\beta_r<0.3$ in red, and $0.4 <\beta_r<0.5$ in blue. The $\beta_r<0.1$  distribution features peak structures on both the near and away side, consistent with di-jet events. One finds however the introduction of even modest boost ($0.2<\beta_r<0.3$) produces a flattening of the away side jet peak, whereas stronger velocities result in a wide dip on the away-side as observed in STAR and PHENIX data.

The above toy model is rather primitive, so we turn to PYTHIA to provide more realistic jets, and underlying events. PYTHIA offers the advantage of providing well tested descriptions of $p+p$ events and jets. It includes effects of momentum conservation, and resonance decay known to influence the shape and strength of correlation functions. A key question is whether, or how, jets couple to radial flow. In this work, we will assume maximum coupling: Jets produced by PYTHIA will be used, without quenching or modification,  to model the effect of radial flow on the shape and strength of 
two- and three-particle correlations. 

This work is divided as follows. Section 2 presents definitions 
and examples of two- and three-particle correlation functions 
based on PYTHIA $p+p$ collisions generated at $\sqrt{s}=200$ GeV. 
In Section 3, we introduce transverse radial flow and show by 
comparison to reference plots of Section 2 how even modest flow 
velocities imparted to PYTHIA events can qualitatively reproduce 
the away-side dip and near-side ridge structures observed experimentally.

\section{Reference Correlations}

\begin{figure}[htb]      
\mbox{
\begin{minipage}{0.25\linewidth} \begin{center}
\includegraphics[height=4.0in,width=1.1\linewidth]{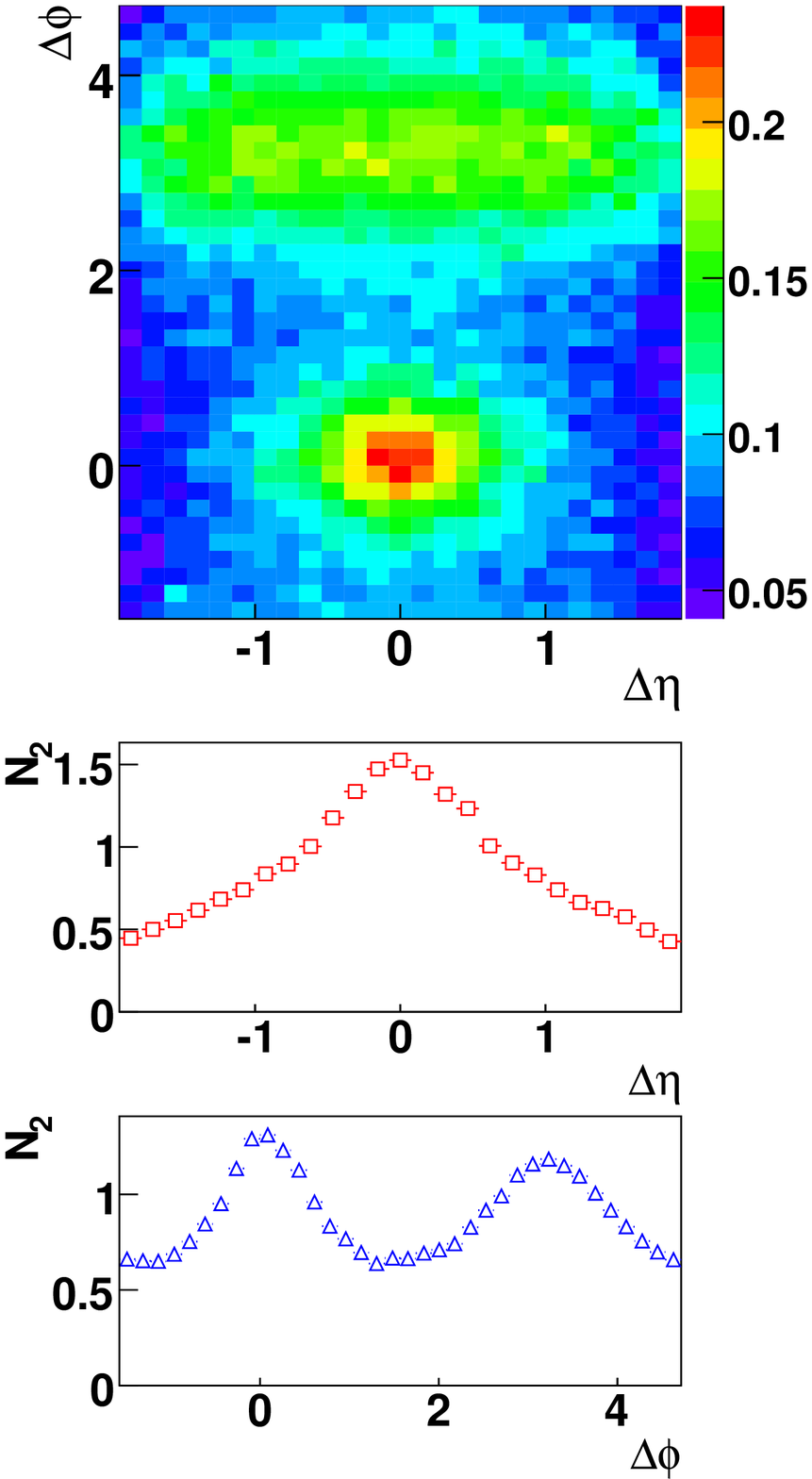}
\end{center}(a) \end{minipage}
\begin{minipage}{0.25\linewidth}\begin{center}
\includegraphics[height=4.0in,width=1.1\linewidth]{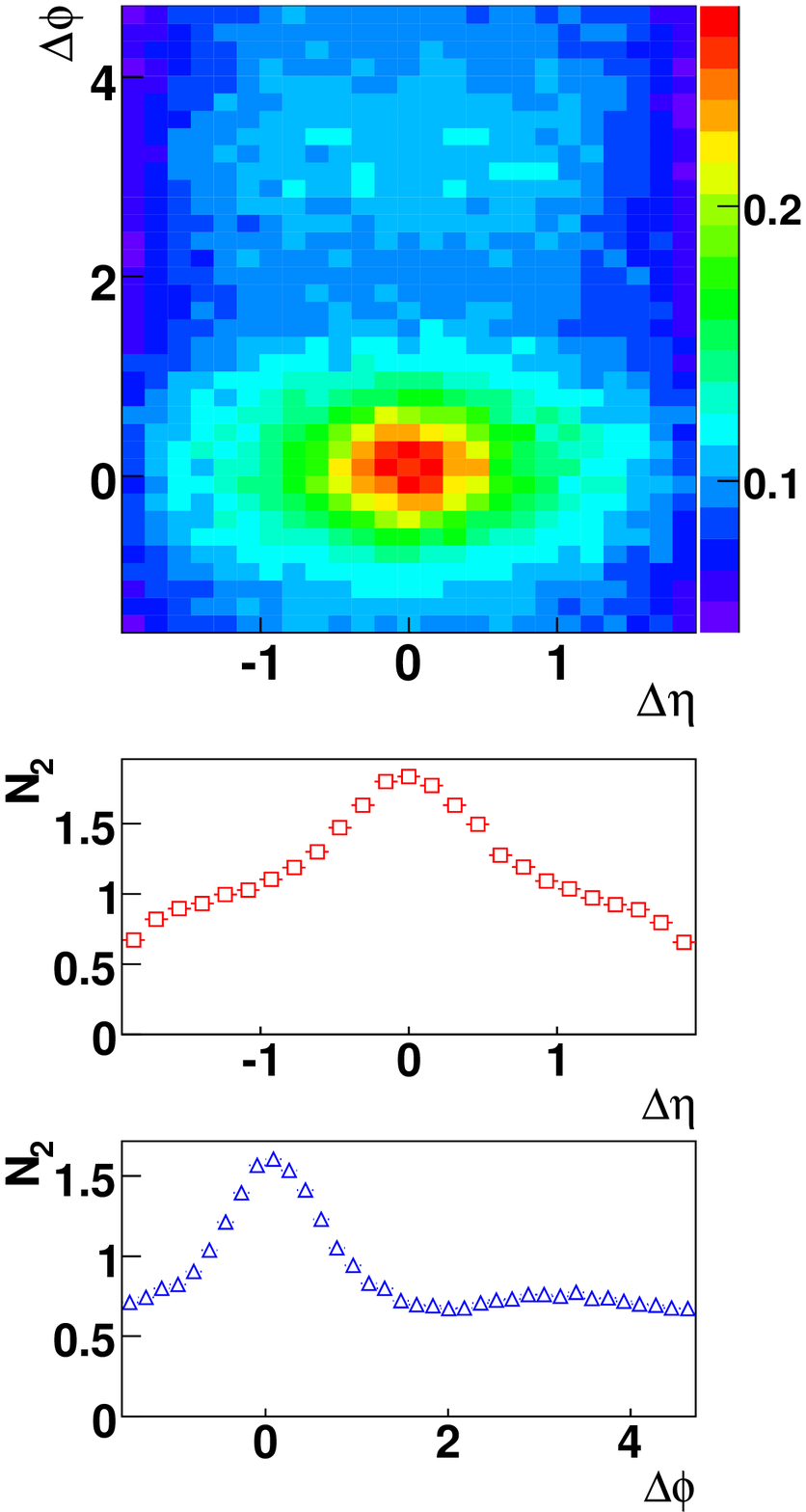}
\end{center}(b)\end{minipage}
\begin{minipage}{0.25\linewidth}\begin{center}
\includegraphics[height=4.0in,width=1.1\linewidth]{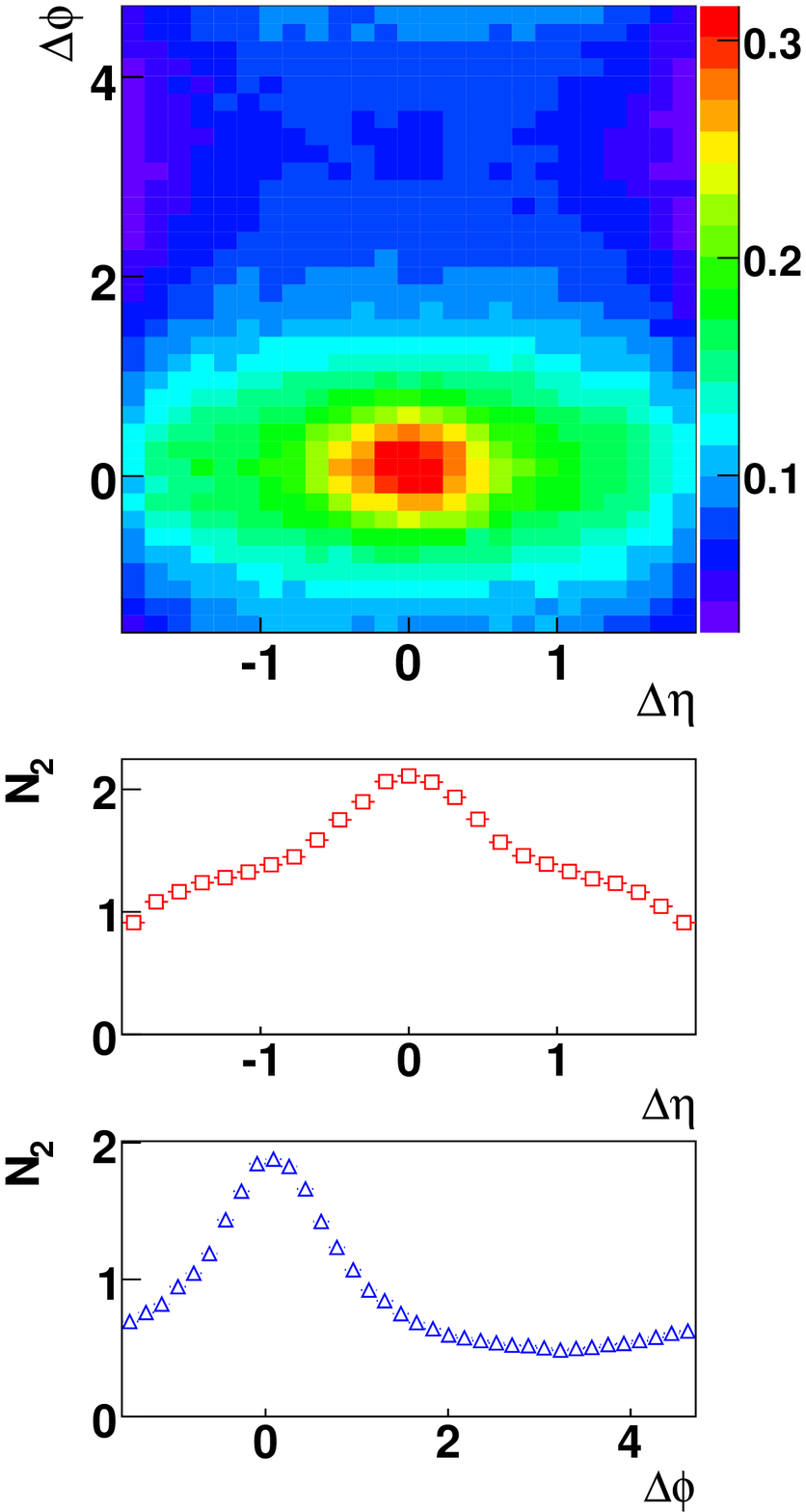}
\end{center}(c) \end{minipage}
\begin{minipage}{0.25\linewidth}\begin{center}
\includegraphics[height=4.0in,width=1.1\linewidth]{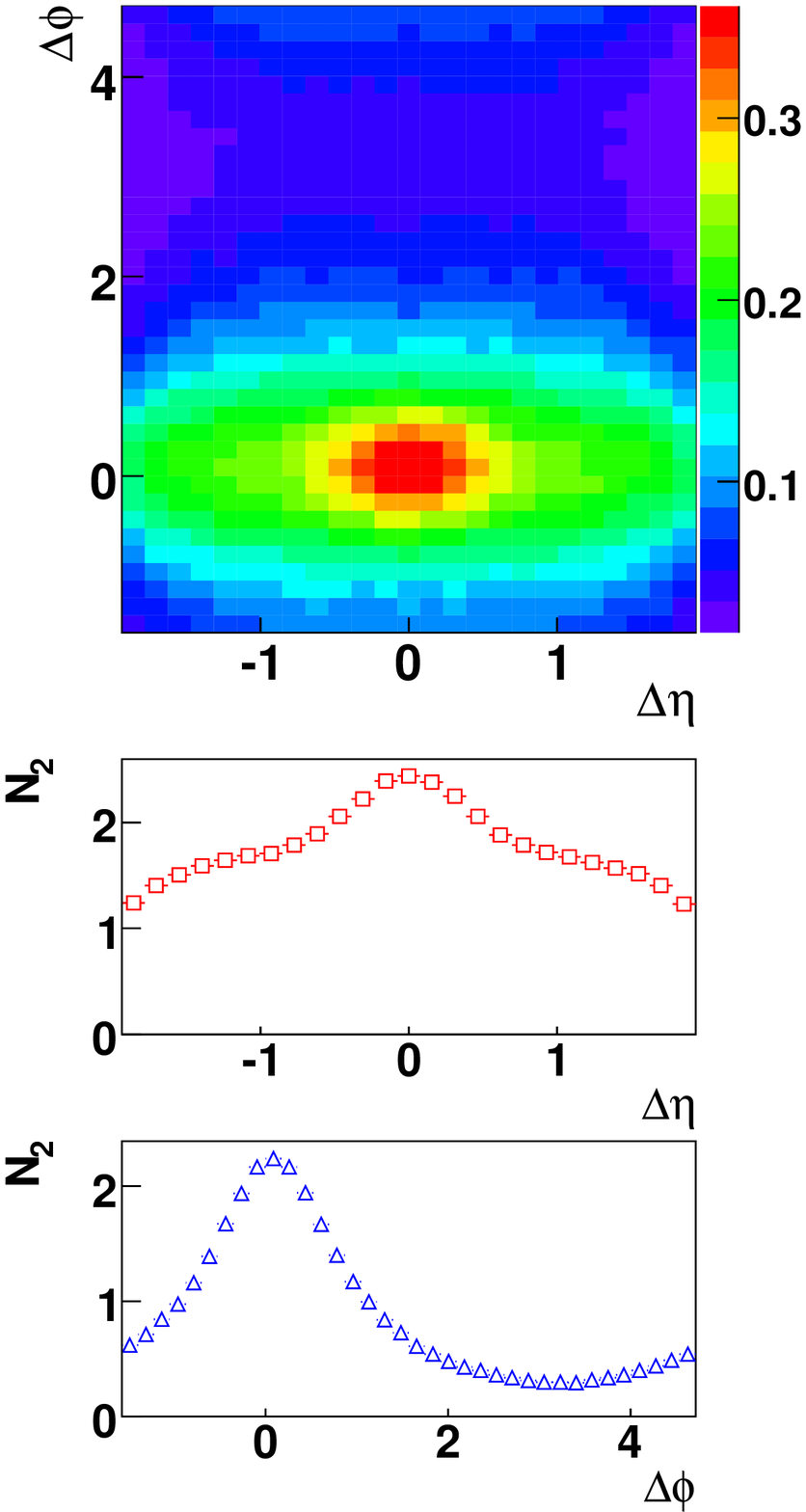}
\end{center}(d) \end{minipage}
}
\caption[]{(Color online) Normalized two-particle 
density, $1/N_{2}^{tot}d^2N_{2}/d\Delta\eta d\Delta\phi$ obtained 
with $p+p$ PYTHIA events:  Particle 1 (jet tag particle) in the 
range $3 < p_{t} < 20$~GeV/c, particles 2 and 3 (associates) in the range 
$0.2 < p_{t} < 1$~GeV/c. Panel (a) through (d) obtained 
with radial boosts $\beta_r$=0.0, 0.2, 0.3, 0.4 respectively.
Middle and bottom subpanels show projections along $\Delta\eta$ for $|\Delta\phi|<0.7$ rad, and
$\Delta\phi$ for $|\Delta\eta|<1.0$. See text 
for details.}
\label{fig:3}
\end{figure}

\begin{figure}[htb]
\mbox{
\begin{minipage}{0.25\linewidth} \begin{center}
\includegraphics[height=4.0in,width=1.1\linewidth]{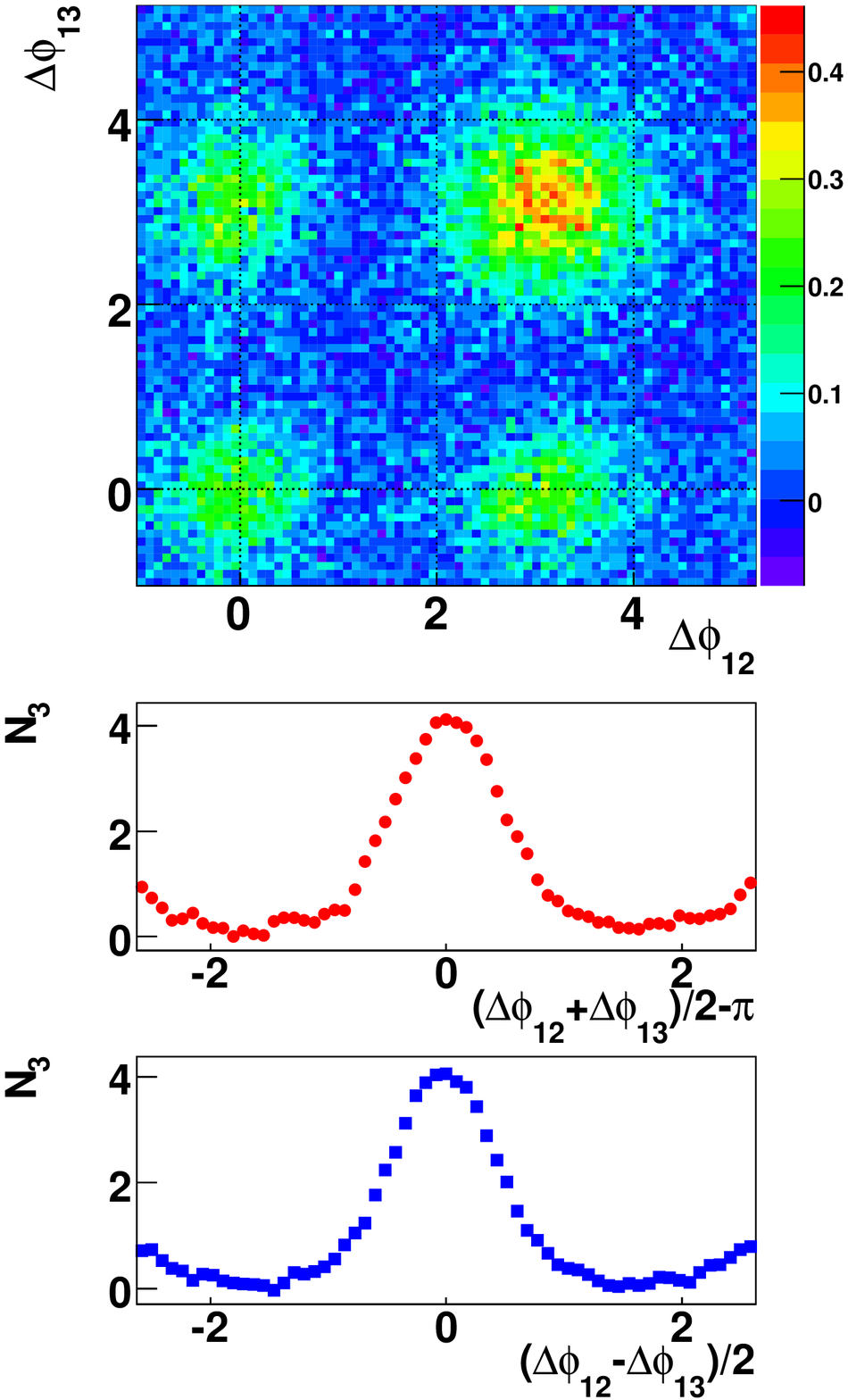}
\end{center}(a)
 \end{minipage}
\begin{minipage}{0.25\linewidth}\begin{center}
\includegraphics[height=4.0in,width=1.1\linewidth]{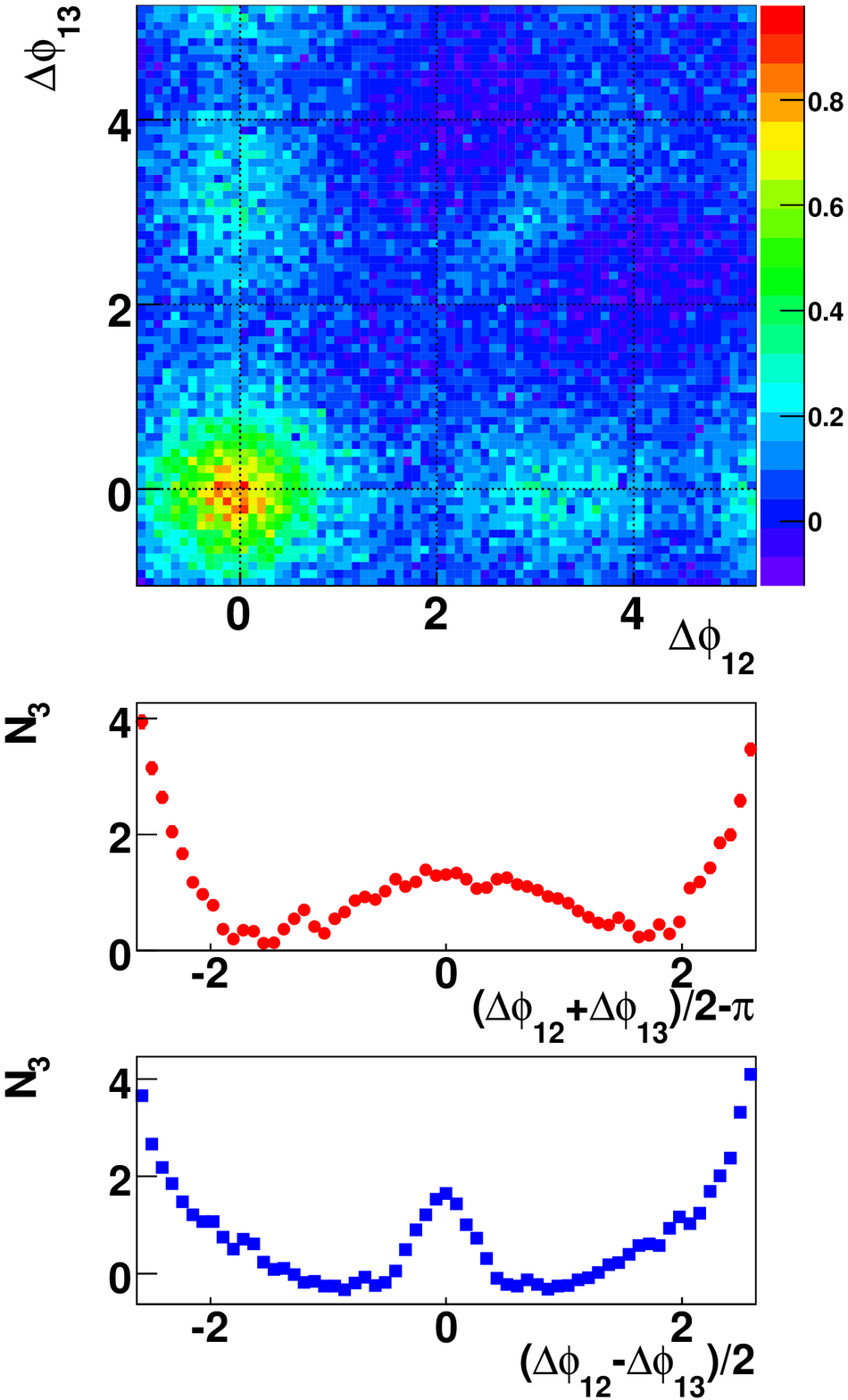}
\end{center}(b)\end{minipage}
\begin{minipage}{0.25\linewidth}\begin{center}
\includegraphics[height=4.0in,width=1.1\linewidth]{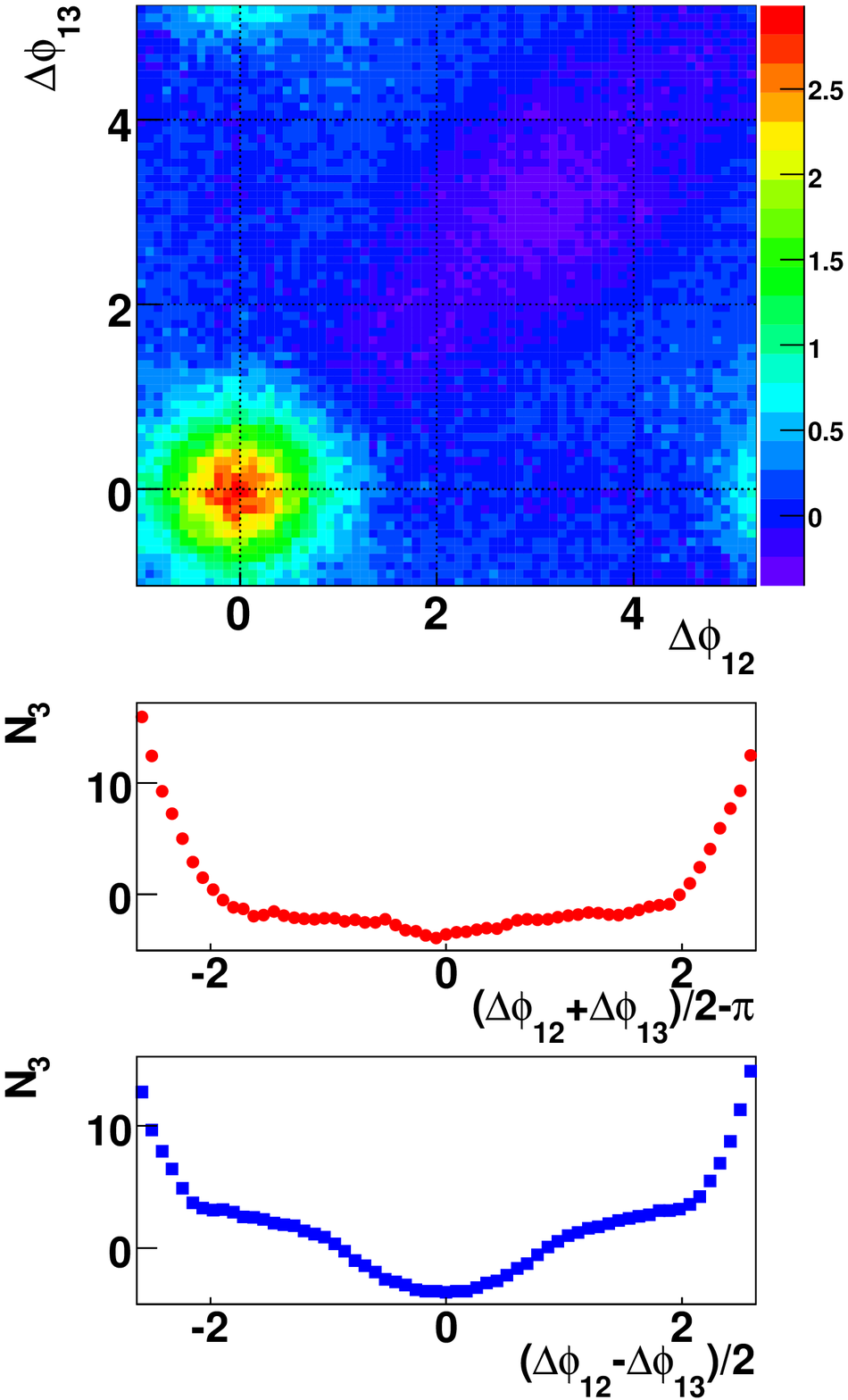}
\end{center}(c)\end{minipage}
\begin{minipage}{0.25\linewidth}\begin{center}
\includegraphics[height=4.0in,width=1.1\linewidth]{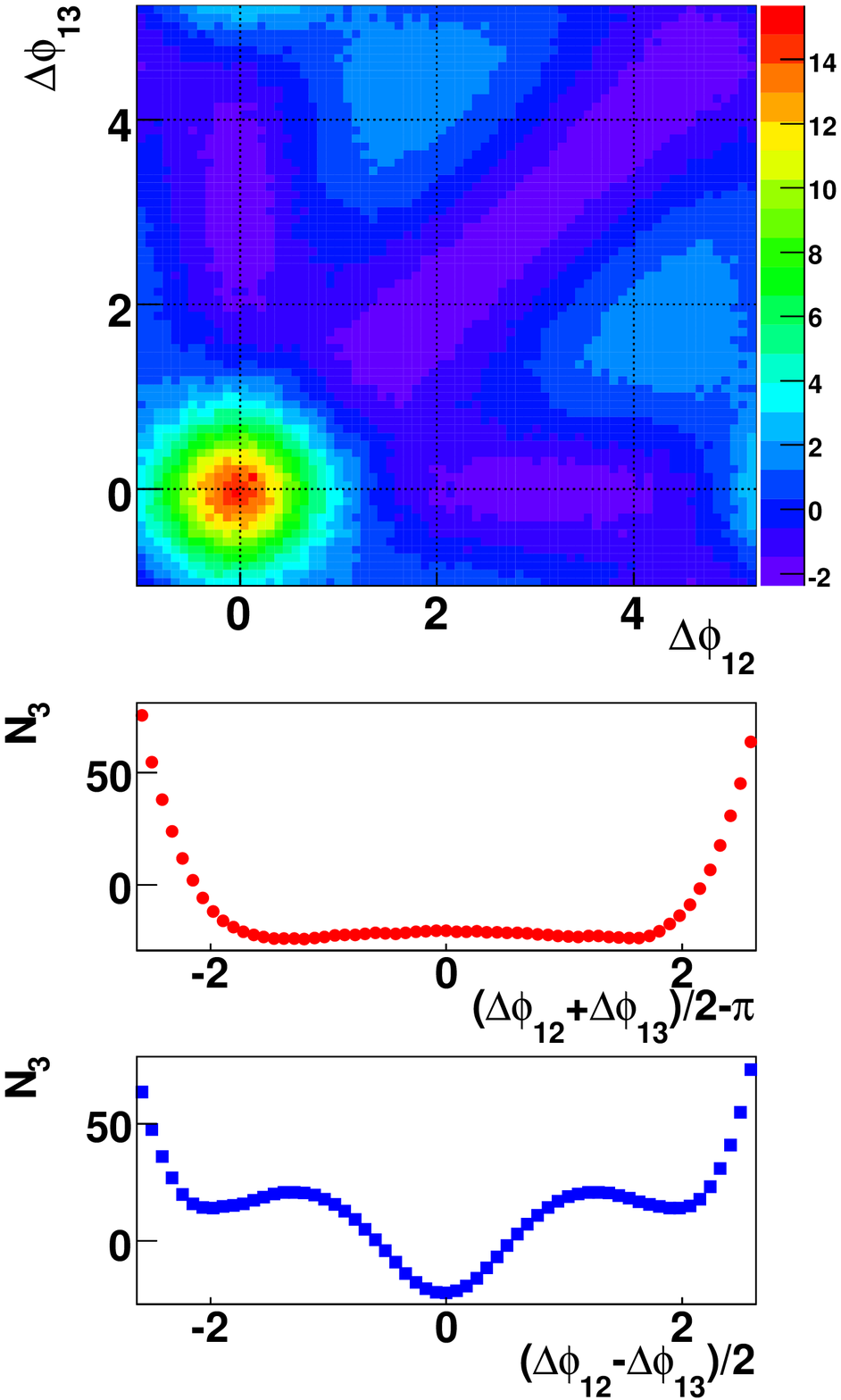}
\end{center}(d)\end{minipage}

}
\caption[]{(Color online) Normalized three-particle azimuthal cumulant  
$C_3(\Delta\phi_{12},\Delta\phi_{13})$ obtained with p+p PYTHIA 
events: Particle 1 (jet tag particle) in the range $3 < p_{t} 
< 20$ GeV/c, particles 2 and 3 (associates) in the range 
$0.2 < p_{t} < 1$ GeV/c. Panel (a) through (d) obtained 
with radial boosts $\beta_r$=0.0, 0.2, 0.3, 0.4 respectively.
Middle and bottom subpanels show projections along cumulant diagonals. See text 
for details.}
\label{fig:4}
\end{figure}

\begin{figure}[htb]      
\mbox{
\begin{minipage}{0.25\linewidth} \begin{center}
\includegraphics[height=4.0in,width=1.1\linewidth]{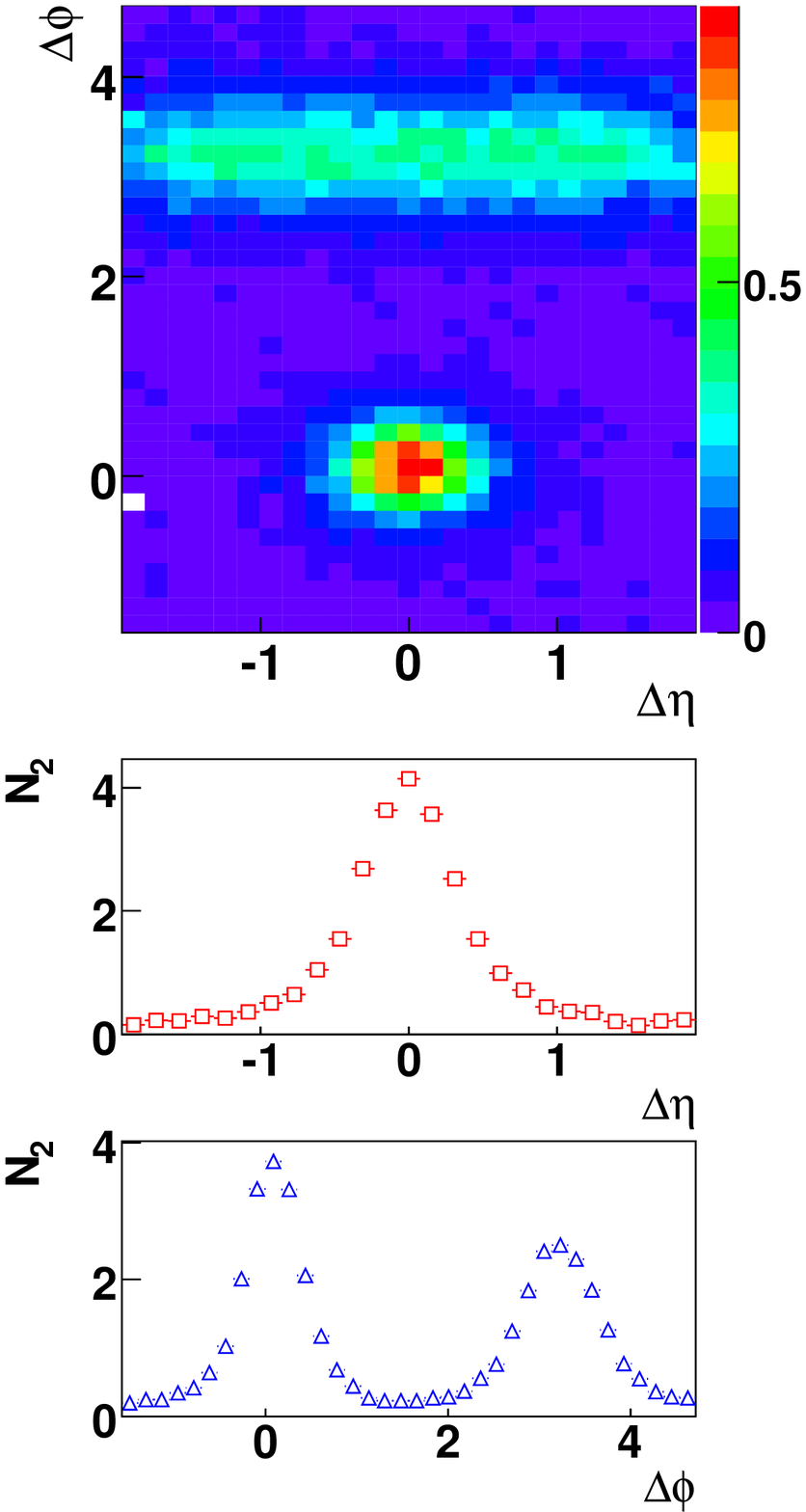}
\end{center}(a) \end{minipage}
\begin{minipage}{0.25\linewidth}\begin{center}
\includegraphics[height=4.0in,width=1.1\linewidth]{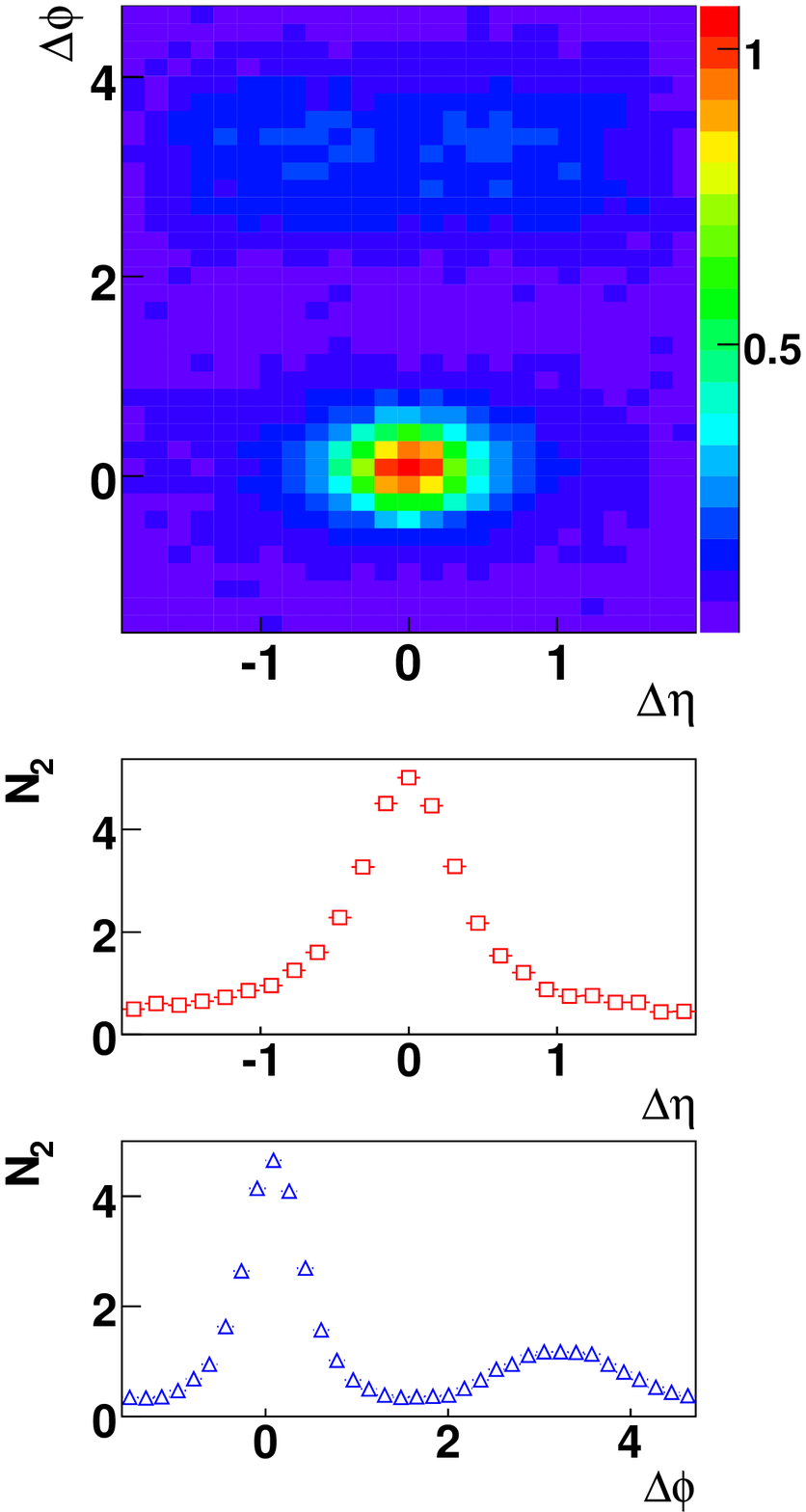}
\end{center}(b)\end{minipage}
\begin{minipage}{0.25\linewidth}\begin{center}
\includegraphics[height=4.0in,width=1.1\linewidth]{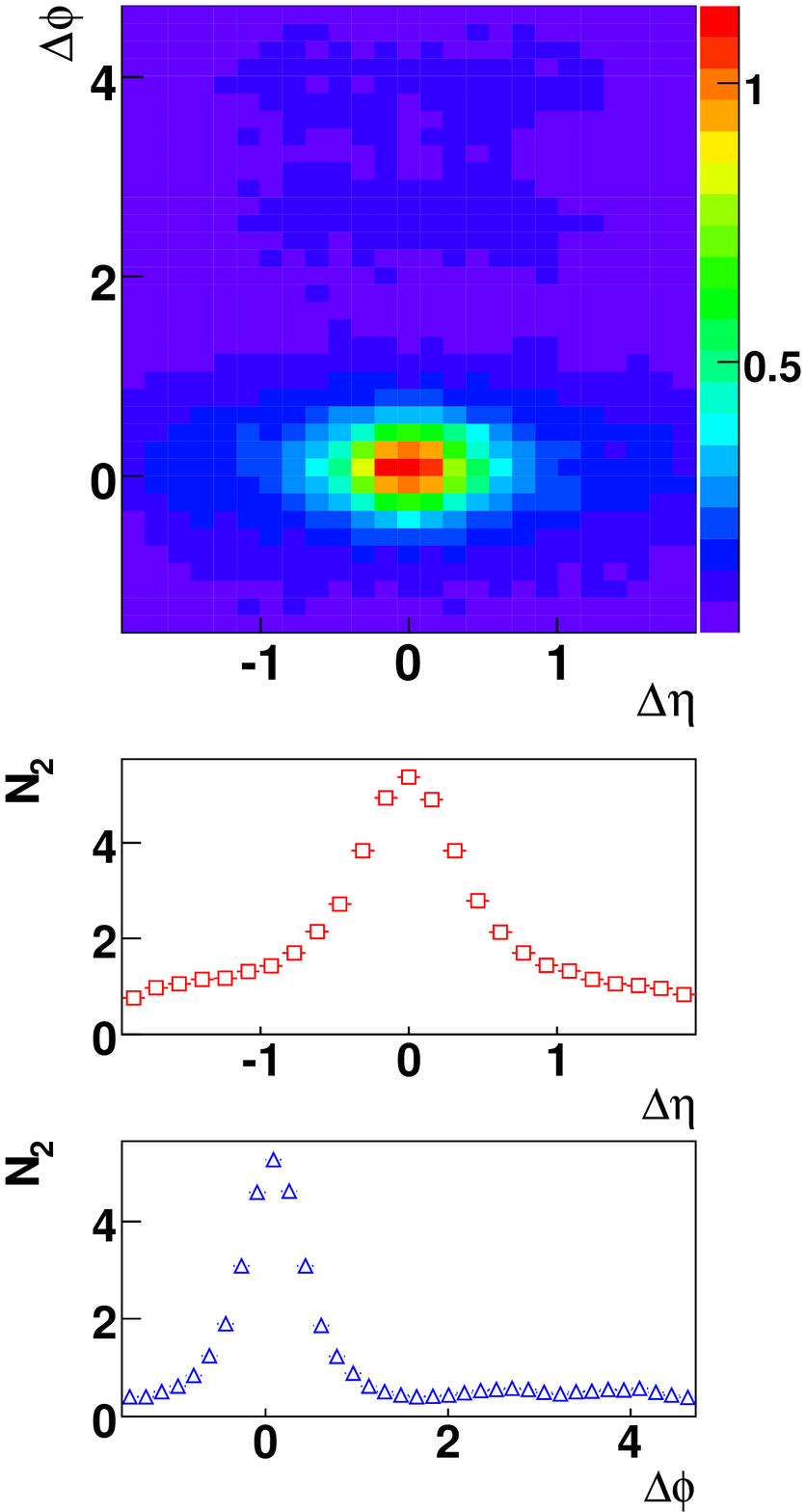}
\end{center}(c) \end{minipage}
\begin{minipage}{0.25\linewidth}\begin{center}
\includegraphics[height=4.0in,width=1.1\linewidth]{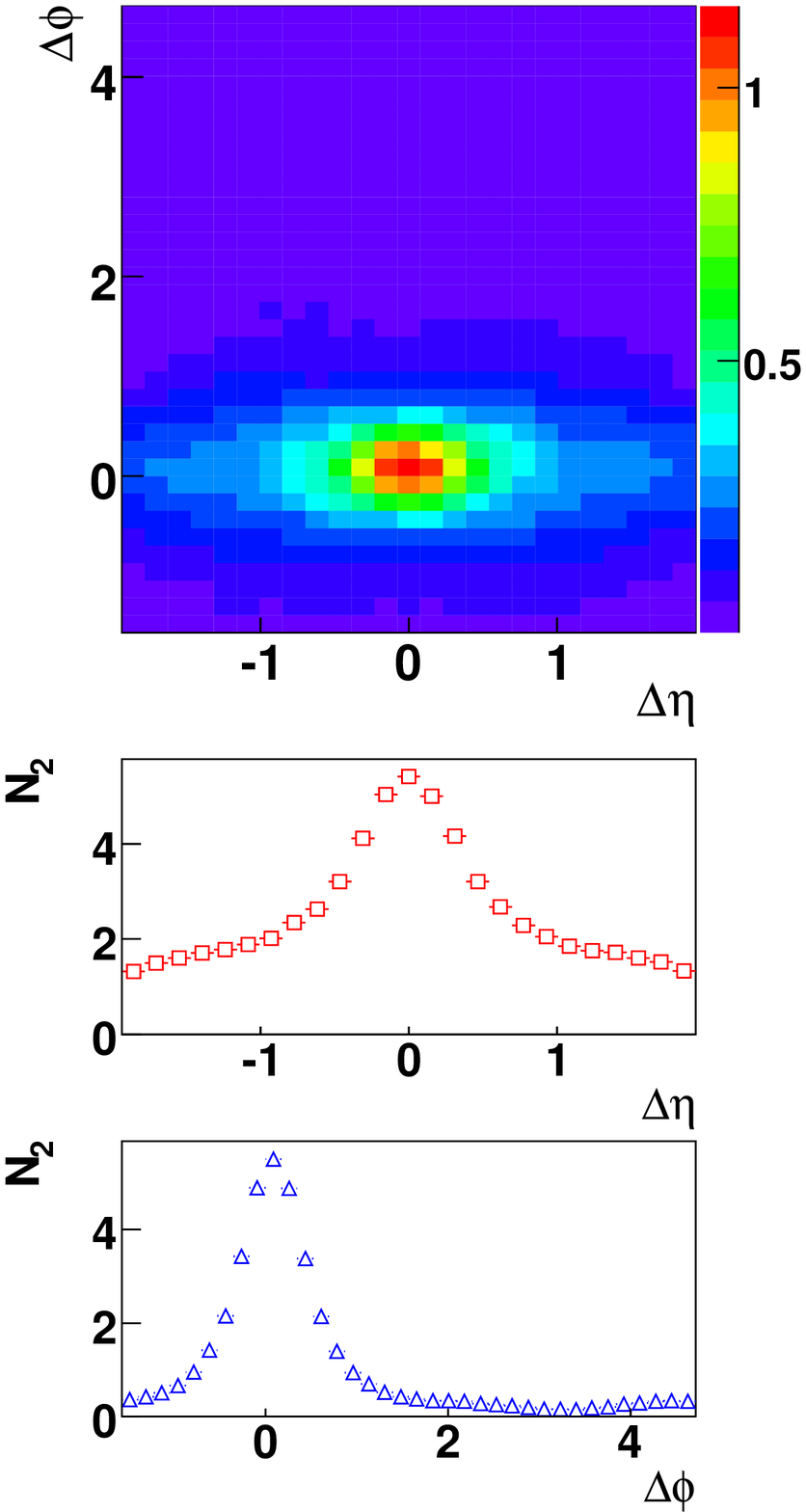}
\end{center}(d) \end{minipage}
}
\caption[]{(Color online)
Normalized two-particle correlations in $\Delta\eta$ vs. $\Delta\phi$  obtained 
with p+p PYTHIA events:  Particle 1 (jet tag particle) in the 
range $3 < p_{t} < 20$~GeV/c, particles 2 and 3 (associates) in the ranges 
$1 < p_{t} < 2$~GeV/c. Panel (a) through (d) obtained 
with radial boosts $\beta_r$=0.0, 0.2, 0.3, 0.4 respectively.
Middle and bottom subpanels show projections along $\Delta\eta$ for $|\Delta\phi|<0.7$ rad, and
$\Delta\phi$ for $|\Delta\eta|<1.0$. See text 
for details.}
\label{fig:5}
\end{figure}

\begin{figure}[htb]
\mbox{
\begin{minipage}{0.25\linewidth} \begin{center}
\includegraphics[height=4.0in,width=1.1\linewidth]{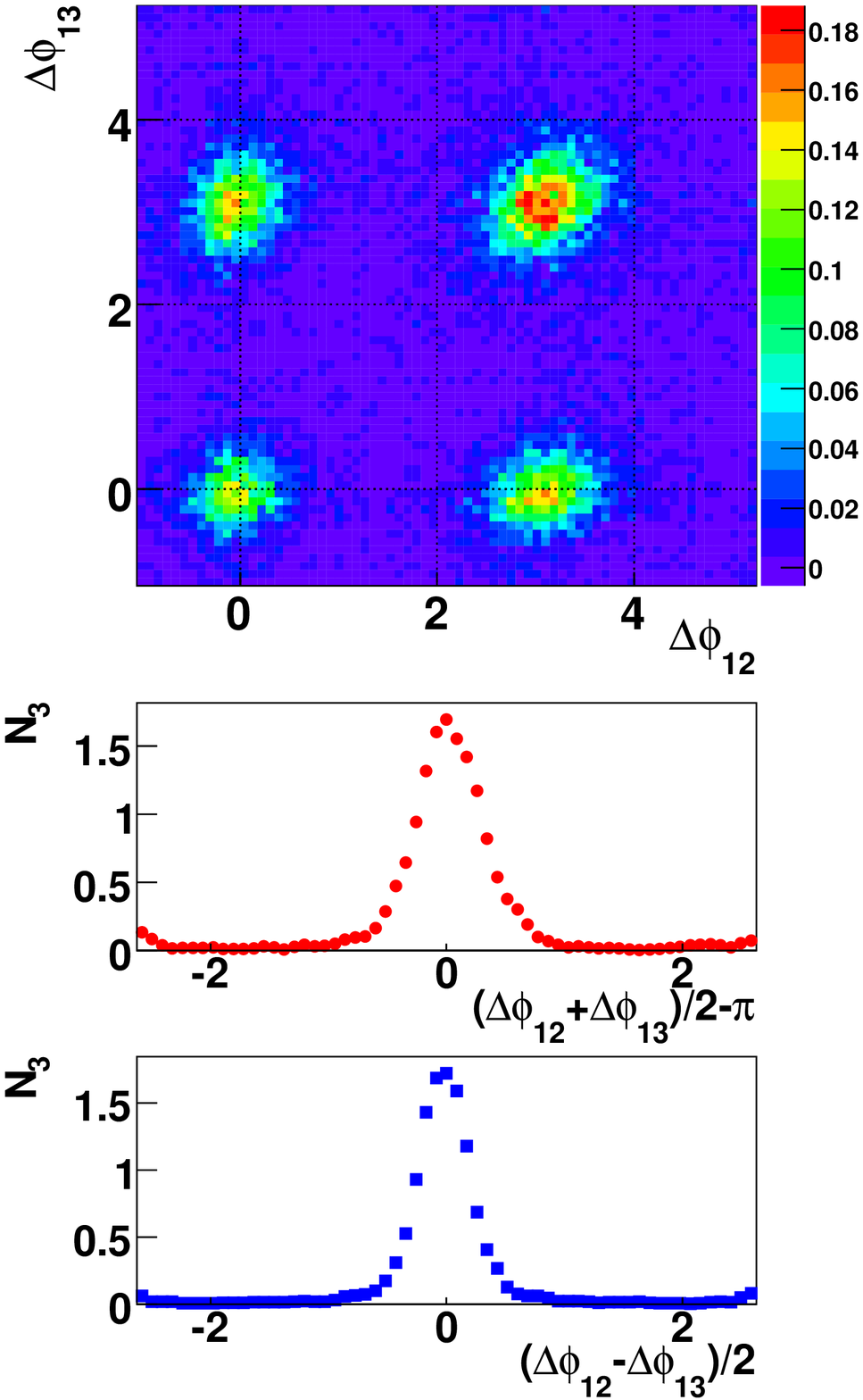}
\end{center}(a)
 \end{minipage}
\begin{minipage}{0.25\linewidth}\begin{center}
\includegraphics[height=4.0in,width=1.1\linewidth]{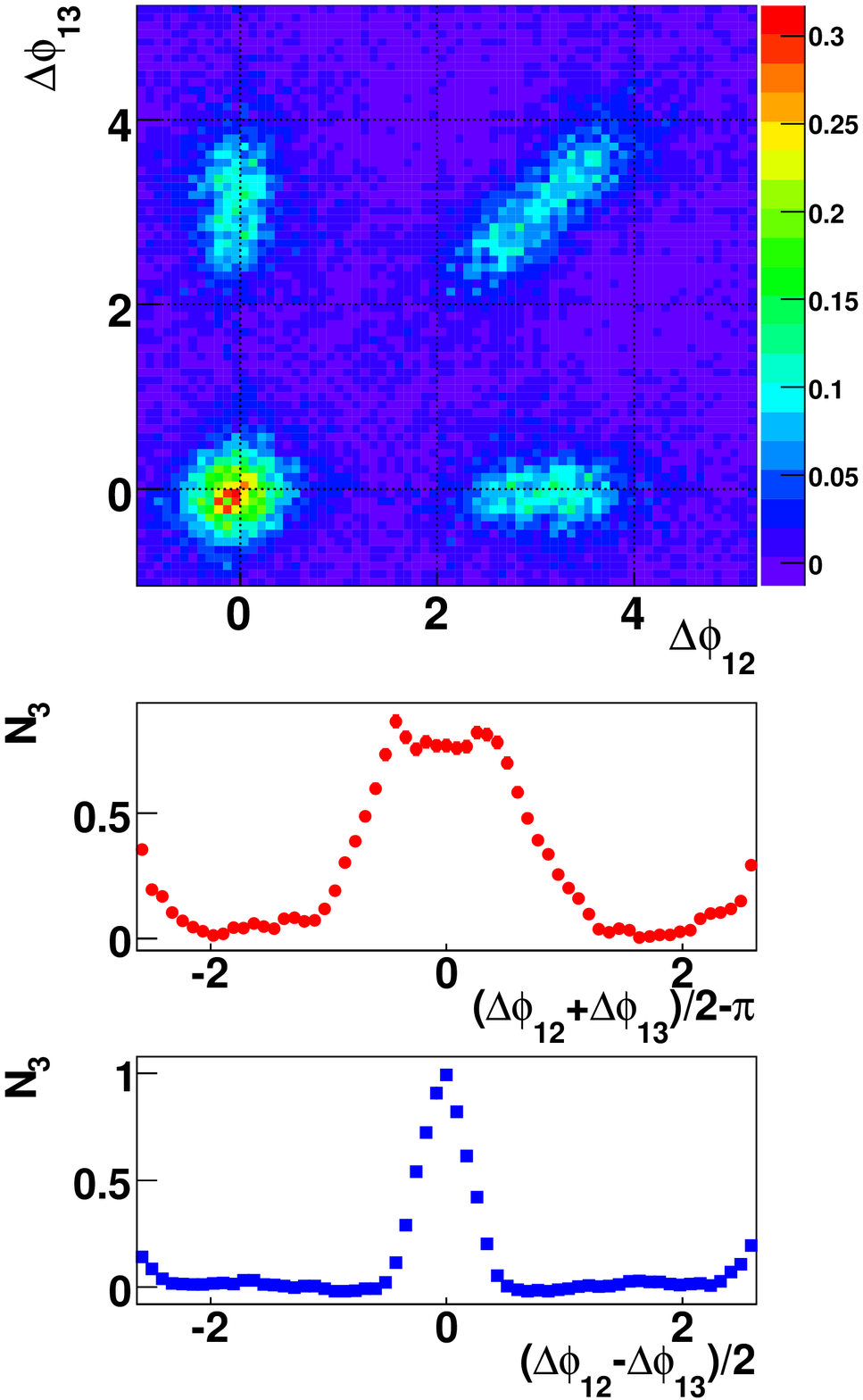}
\end{center}(b)\end{minipage}
\begin{minipage}{0.25\linewidth}\begin{center}
\includegraphics[height=4.0in,width=1.1\linewidth]{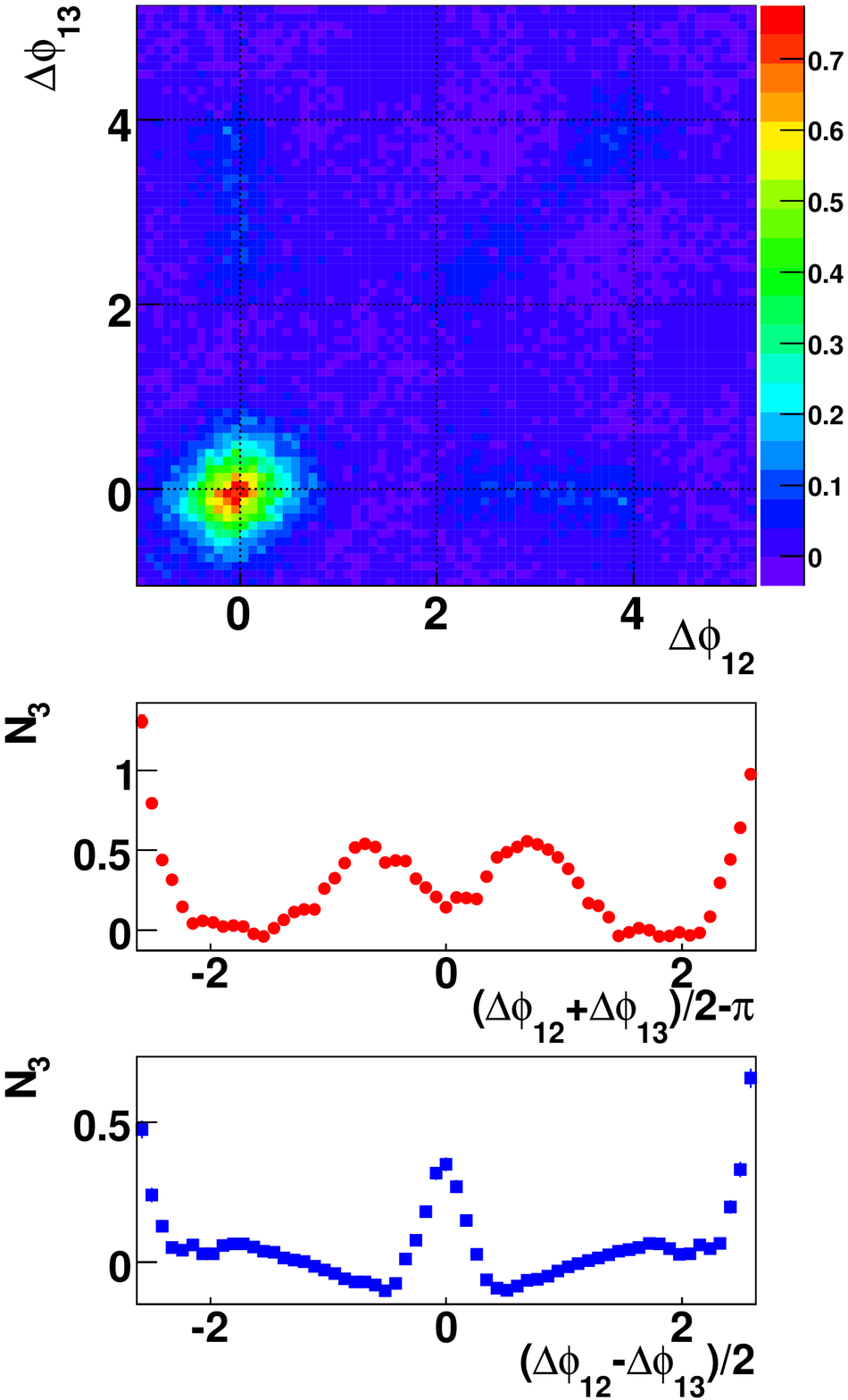}
\end{center}(c)\end{minipage}
\begin{minipage}{0.25\linewidth}\begin{center}
\includegraphics[height=4.0in,width=1.1\linewidth]{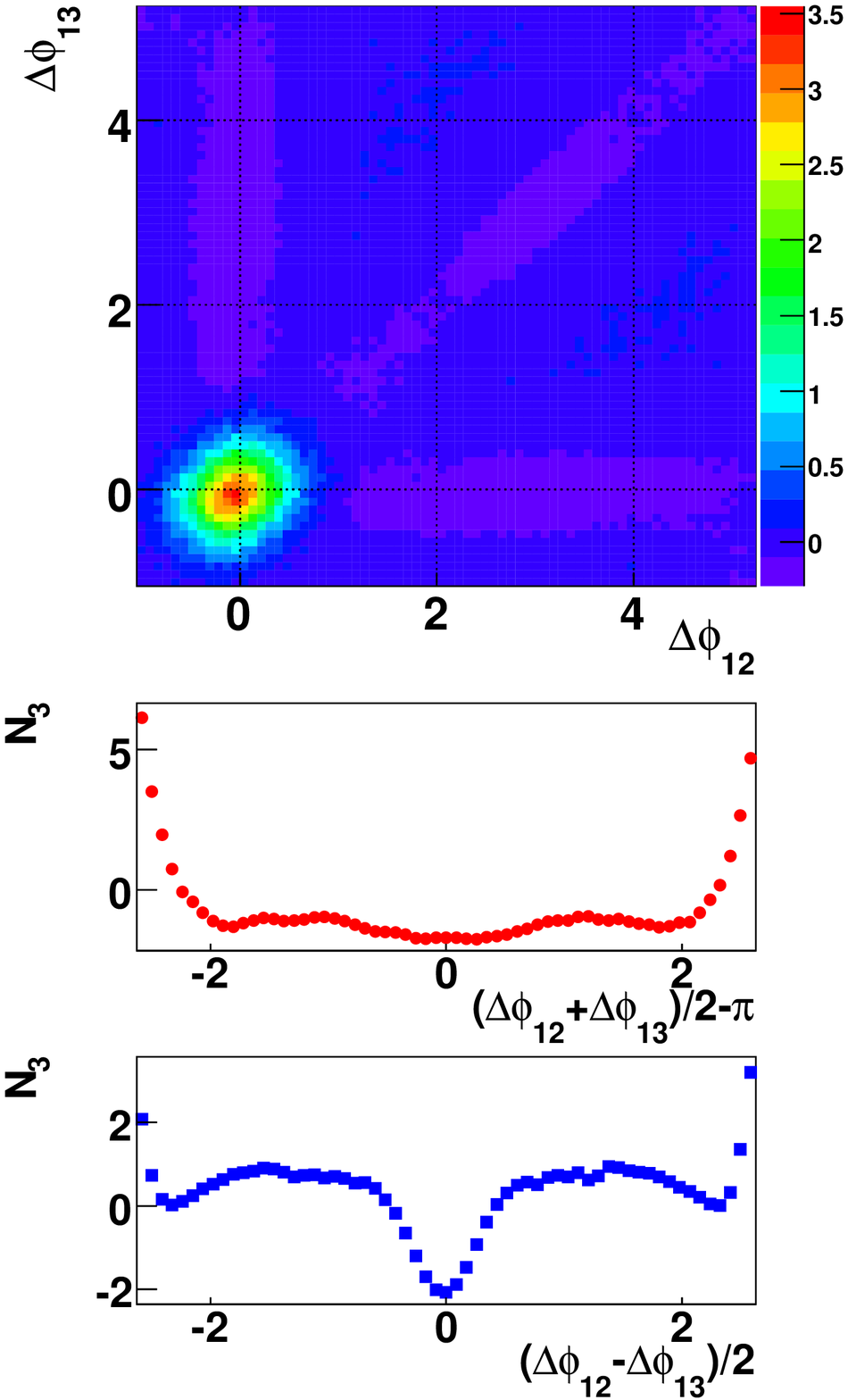}
\end{center}(d)\end{minipage}

}
\caption[]{(Color online) Normalized three-particle azimuthal cumulant  
$C_3(\Delta\phi_{12},\Delta\phi_{13})$ obtained with p+p PYTHIA 
events: Particle 1 (jet tag particle) in the range $3 < p_{t} 
< 20$ GeV/c, particles 2 and 3 (associates) in the range 
$1 < p_{t} < 2$ GeV/c. Panel (a) through (d) obtained 
with radial boosts $\beta_r$=0.0, 0.2, 0.3, 0.4 respectively.
Middle and bottom subpanels show projections along cumulant diagonals. See text 
for details.}
\label{fig:6}
\end{figure}

\begin{figure}[htb]      
\mbox{
\begin{minipage}{0.25\linewidth} \begin{center}
\includegraphics[height=4.0in,width=1.1\linewidth]{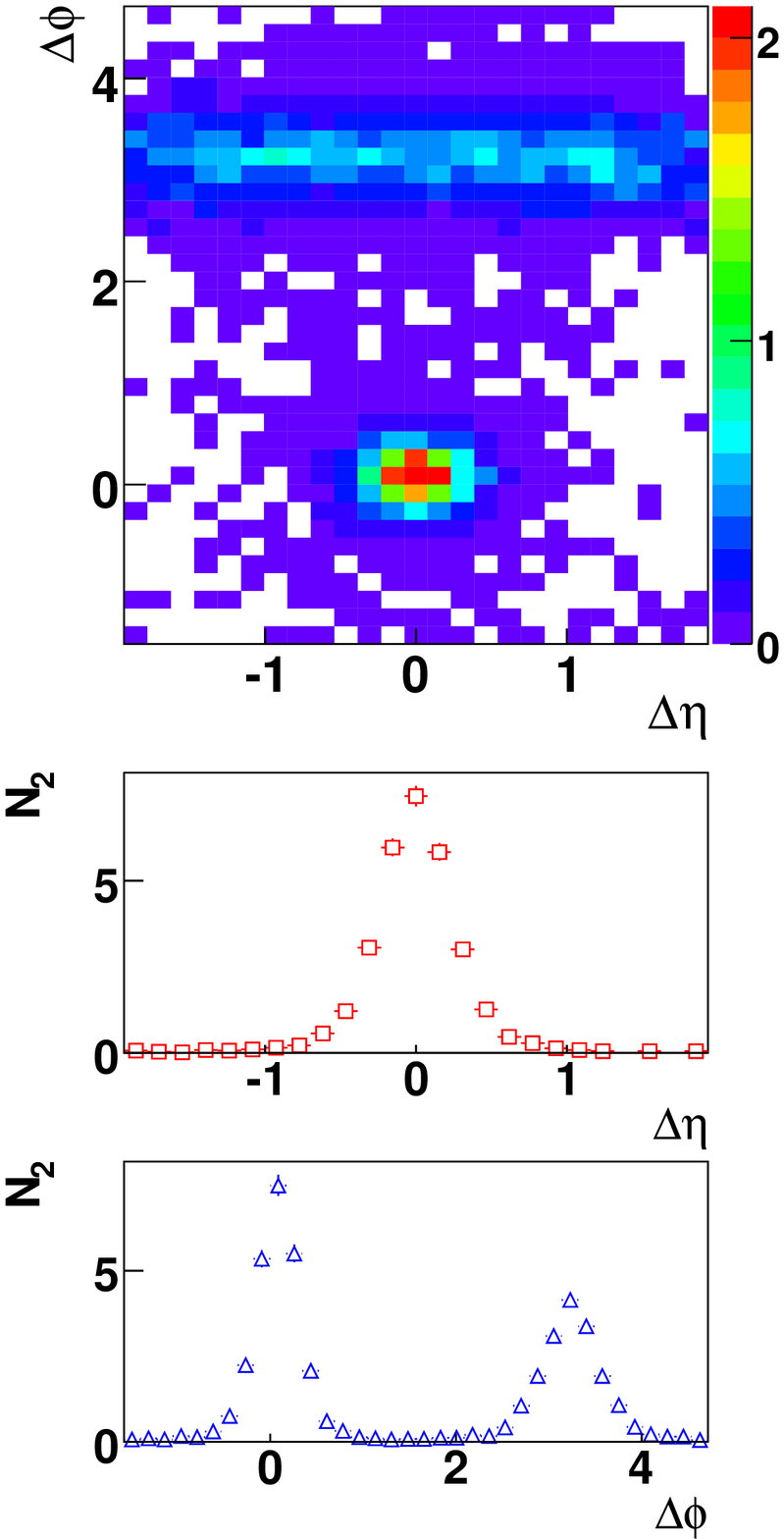}
\end{center}(a) \end{minipage}
\begin{minipage}{0.25\linewidth}\begin{center}
\includegraphics[height=4.0in,width=1.1\linewidth]{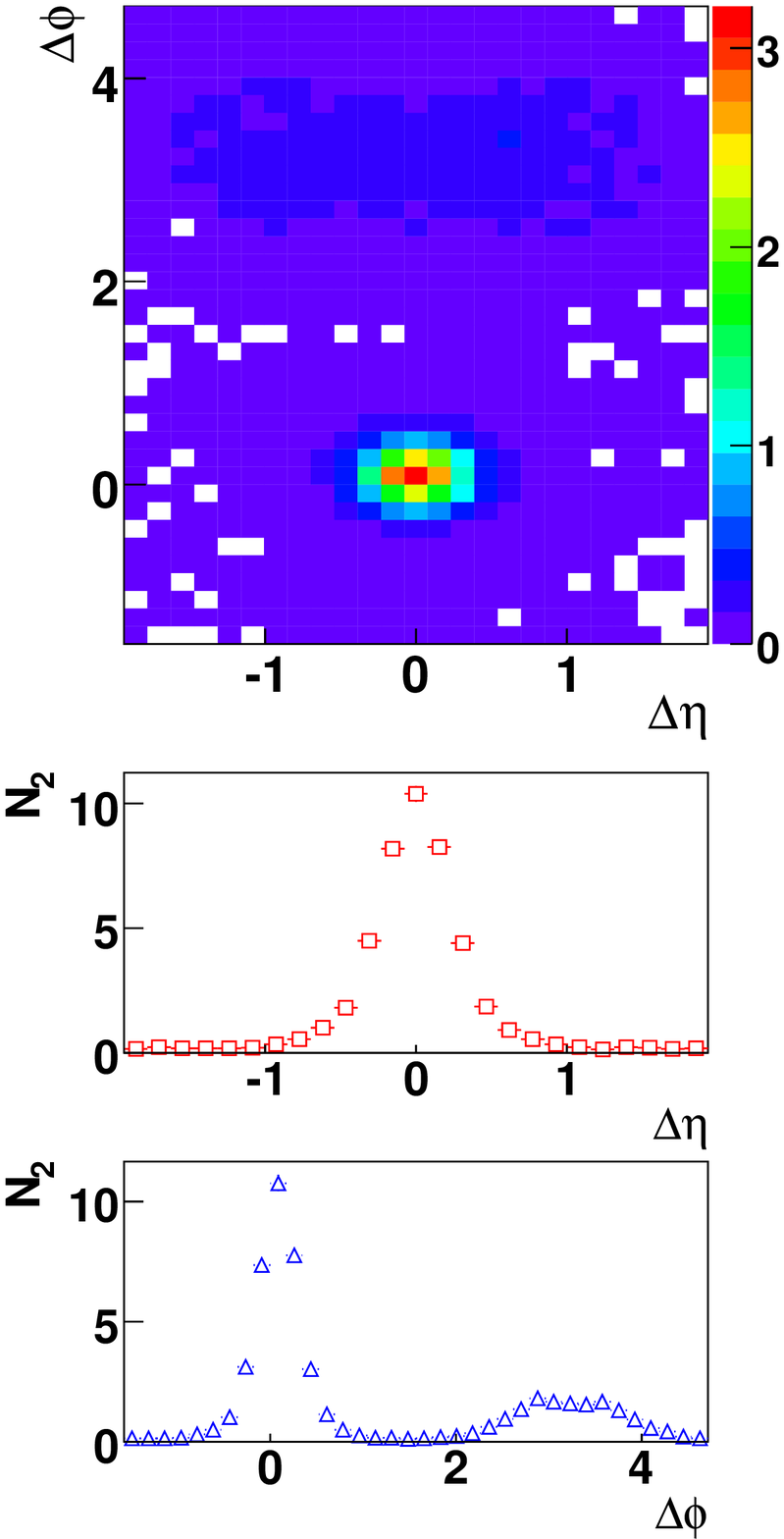}
\end{center}(b)\end{minipage}
\begin{minipage}{0.25\linewidth}\begin{center}
\includegraphics[height=4.0in,width=1.1\linewidth]{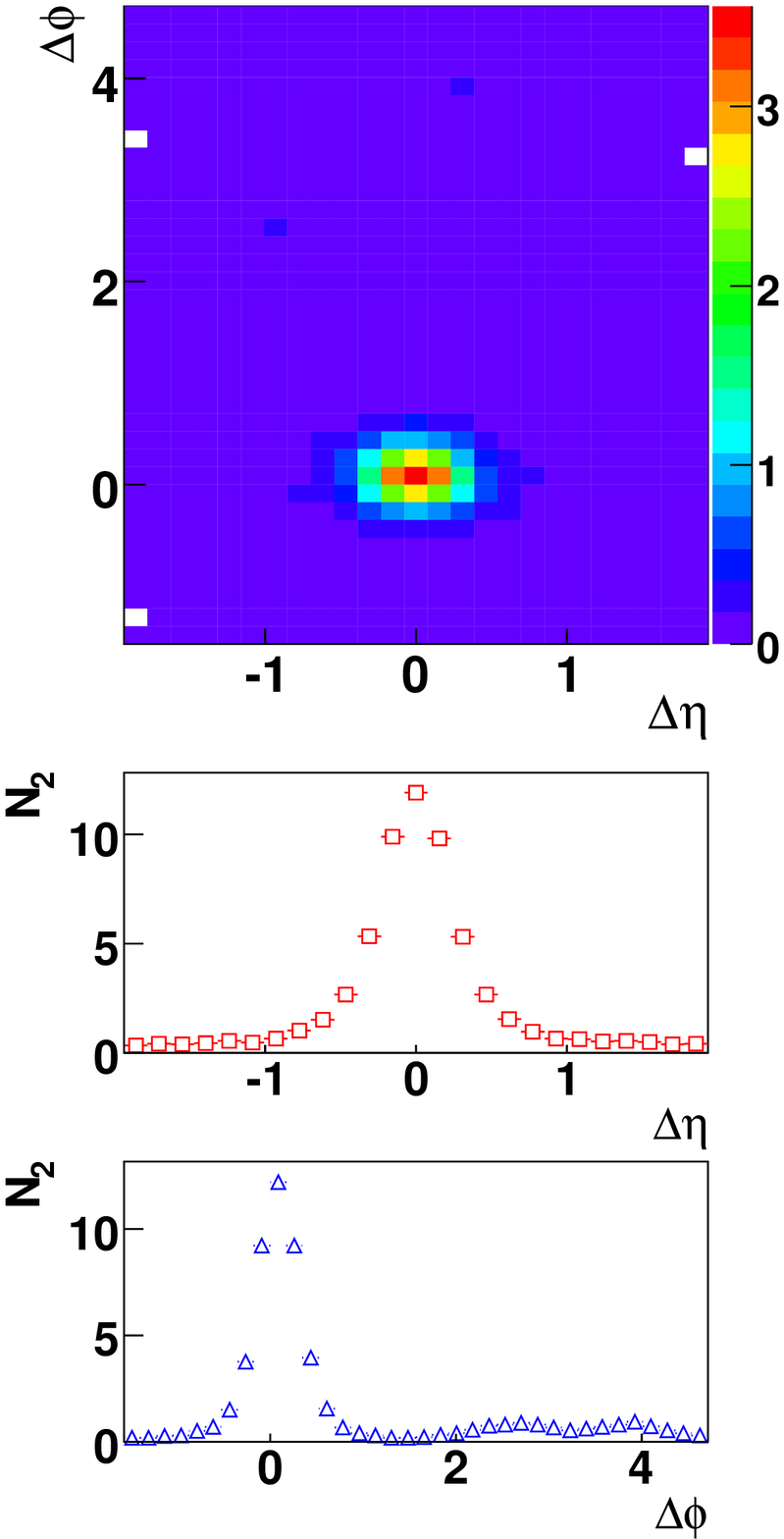}
\end{center}(c) \end{minipage}
\begin{minipage}{0.25\linewidth}\begin{center}
\includegraphics[height=4.0in,width=1.1\linewidth]{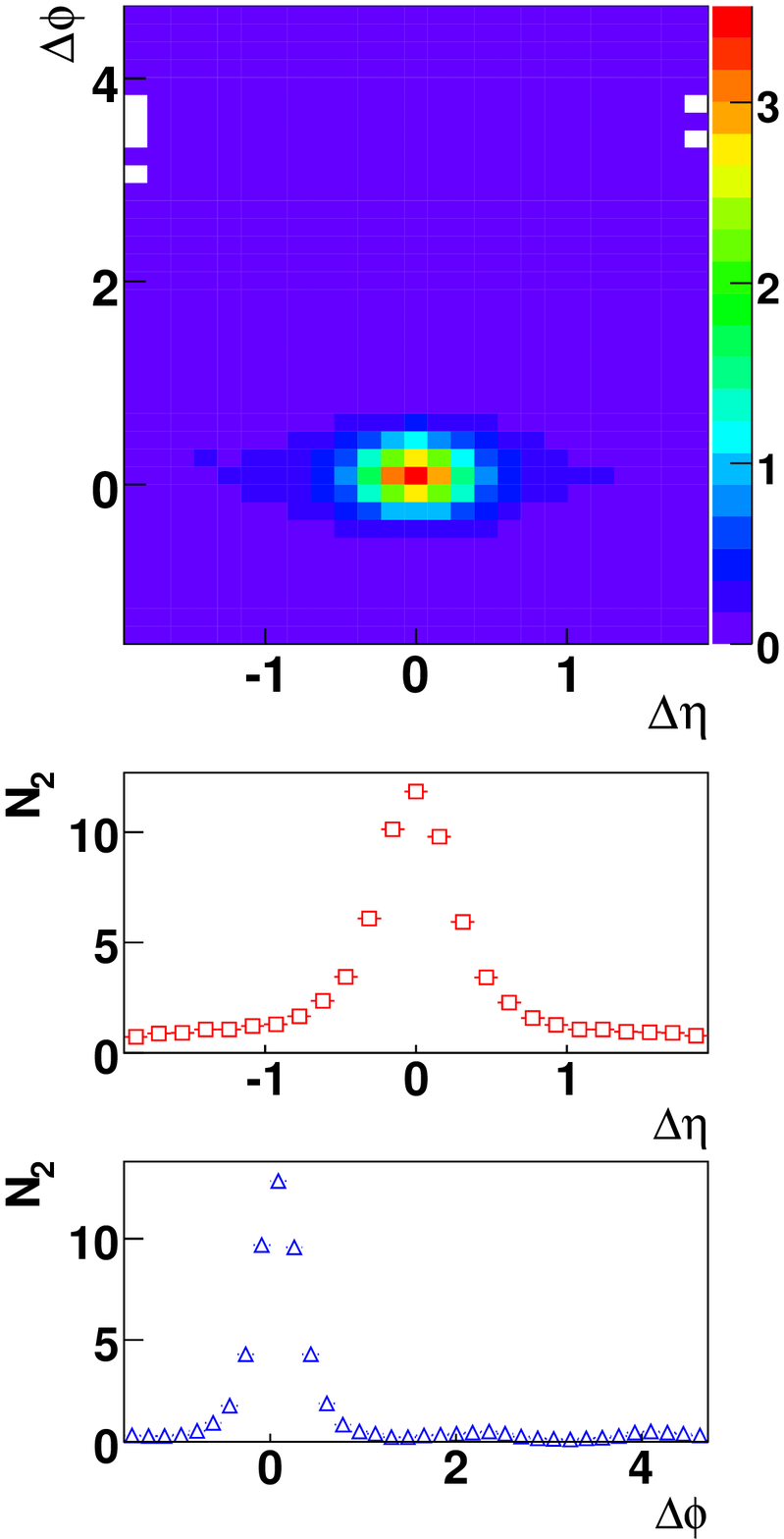}
\end{center}(d) \end{minipage}
}
\caption[]{(Color online)
Normalized two-particle correlations in $\Delta\eta$ vs. $\Delta\phi$  obtained 
with p+p PYTHIA events:  Particle 1 (jet tag particle) in the 
range $3 < p_{t} < 20$~GeV/c, particles 2 and 3 (associates) in the ranges 
$2 < p_{t} < 3$~GeV/c. Panel (a) through (d) obtained 
with radial boosts $\beta_r$=0.0, 0.2, 0.3, 0.4 respectively.
Middle and bottom subpanels show projections along $\Delta\eta$ for $|\Delta\phi|<0.7$ rad, and
$\Delta\phi$ for $|\Delta\eta|<1.0$. See text 
for details.}
\label{fig:7}
\end{figure}

\begin{figure}[htb]
\mbox{
\begin{minipage}{0.25\linewidth} \begin{center}
\includegraphics[height=4.0in,width=1.1\linewidth]{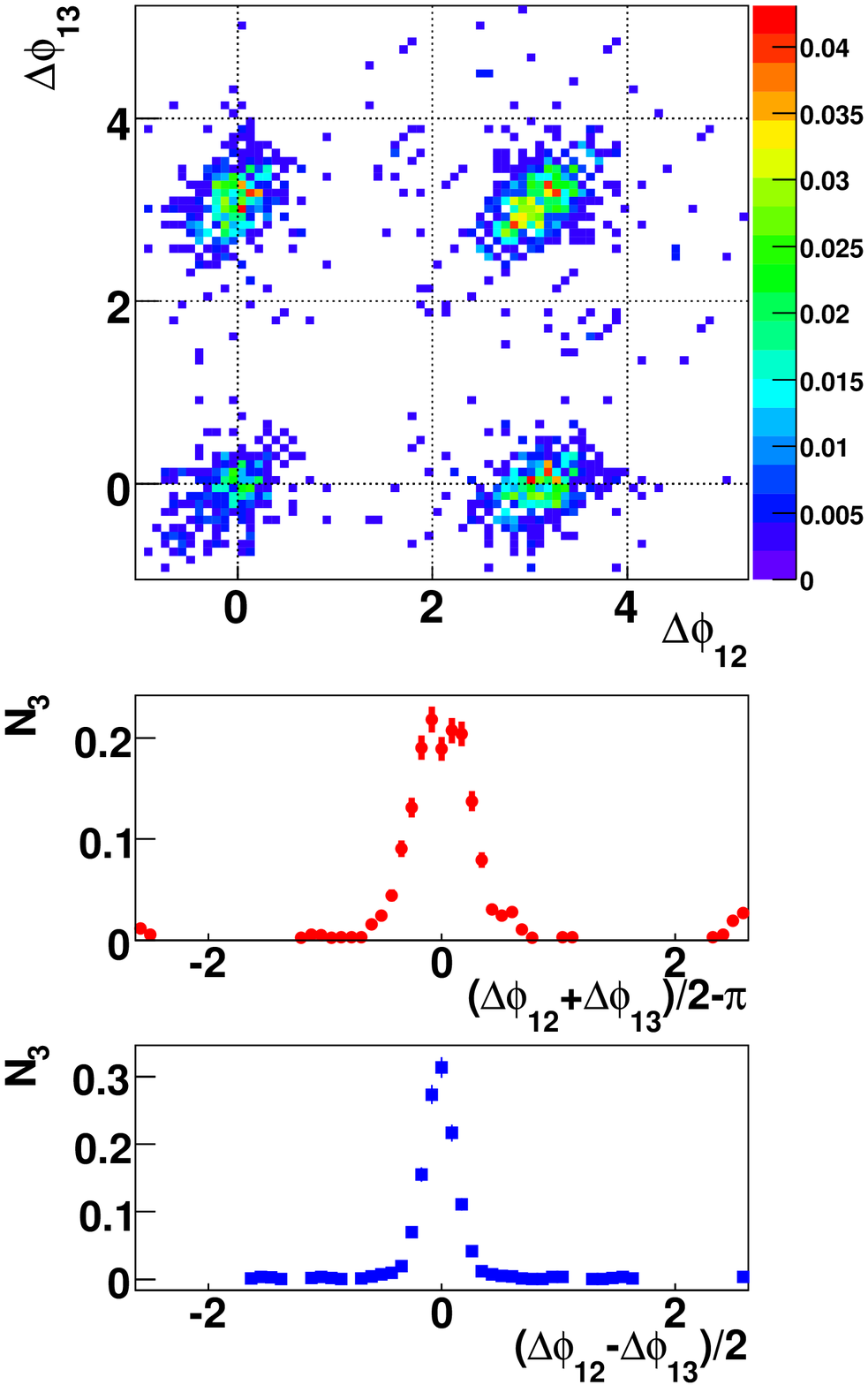}
\end{center}(a)
 \end{minipage}
\begin{minipage}{0.25\linewidth}\begin{center}
\includegraphics[height=4.0in,width=1.1\linewidth]{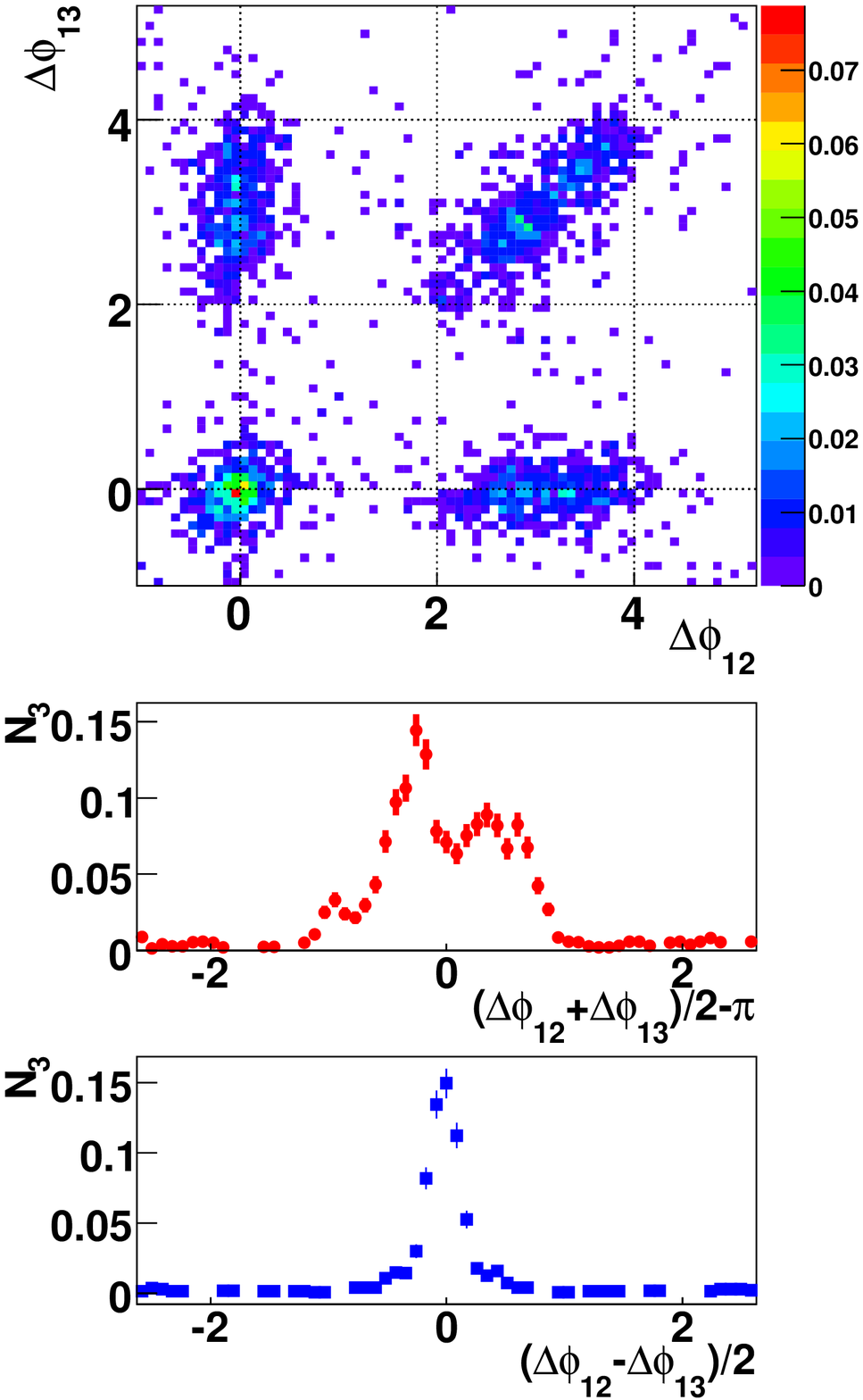}
\end{center}(b)\end{minipage}
\begin{minipage}{0.25\linewidth}\begin{center}
\includegraphics[height=4.0in,width=1.1\linewidth]{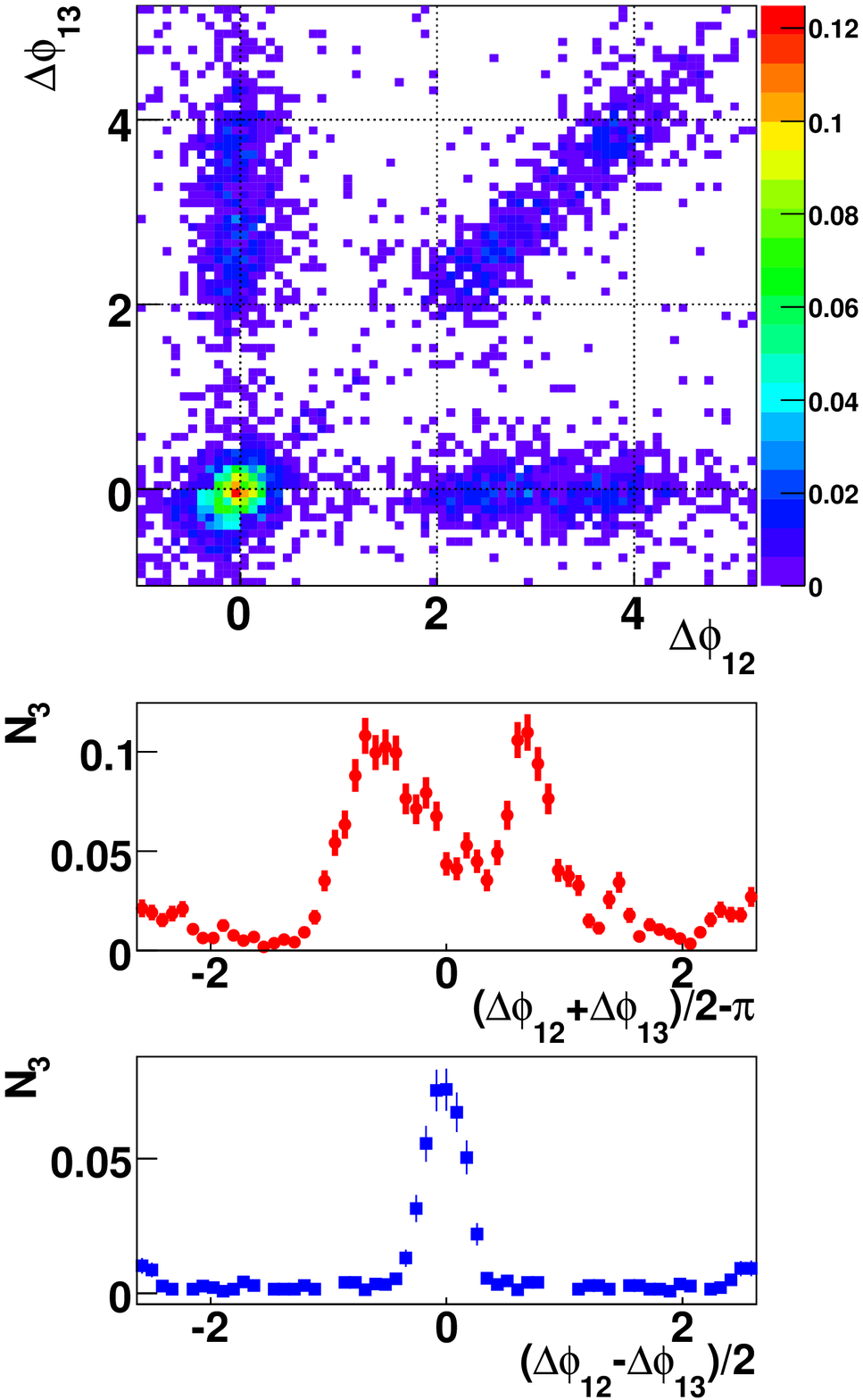}
\end{center}(c)\end{minipage}
\begin{minipage}{0.25\linewidth}\begin{center}
\includegraphics[height=4.0in,width=1.1\linewidth]{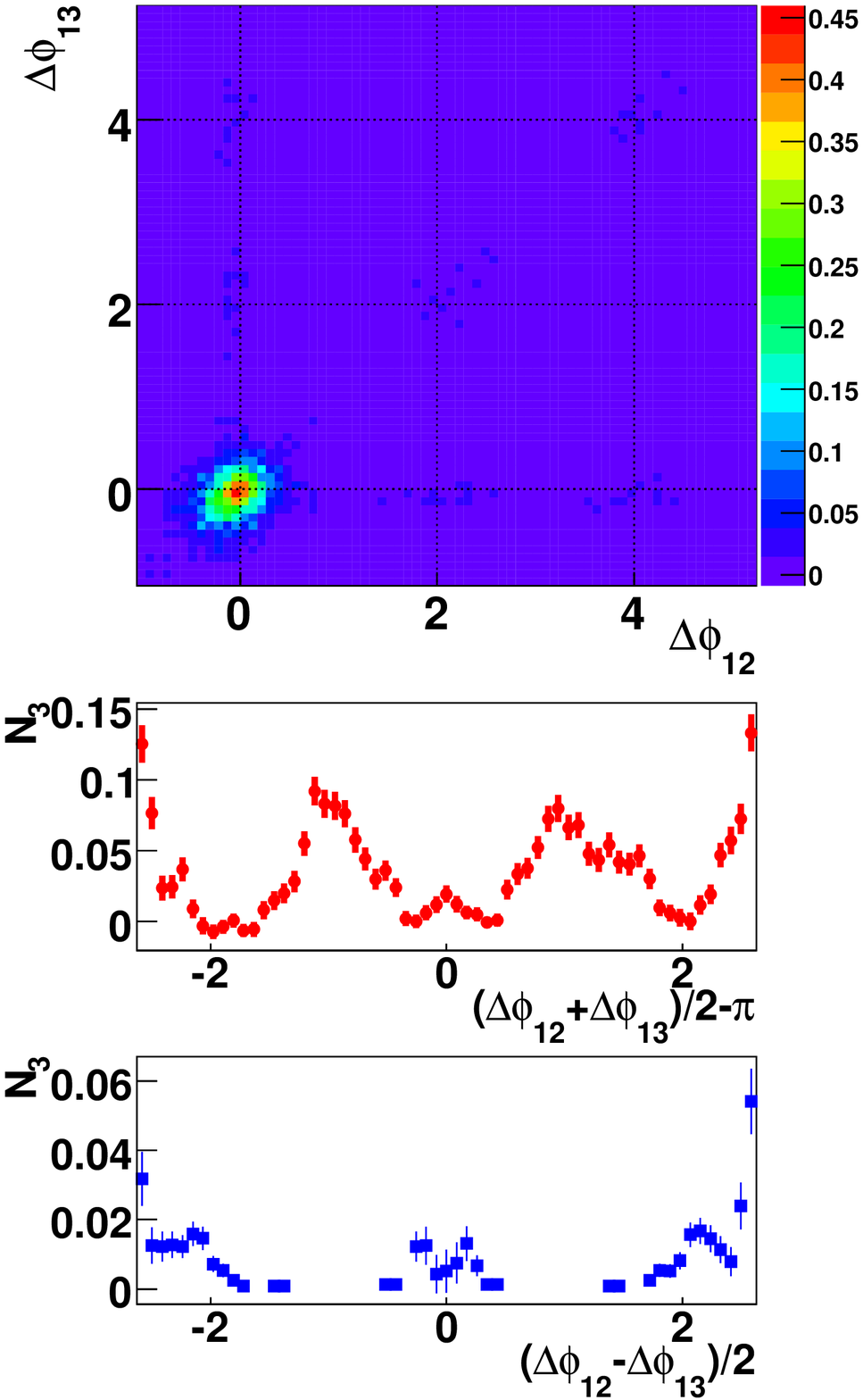}
\end{center}(d)\end{minipage}

}
\caption[]{(Color online) Normalized three-particle azimuthal cumulant  
$C_3(\Delta\phi_{12},\Delta\phi_{13})$ obtained with p+p PYTHIA 
events: Particle 1 (jet tag particle) in the range $3 < p_{t} 
< 20$ GeV/c, particles 2 and 3 (associates) in the range 
$2 < p_{t} < 3$ GeV/c. Panel (a) through (d) obtained 
with radial boosts $\beta_r$=0.0, 0.2, 0.3, 0.4 respectively.
Middle and bottom panels show projections along cumulant diagonals. See text 
for details.}
\label{fig:8}
\end{figure}

We first present reference correlations based on p+p collisions 
at $\sqrt{s}=200$~GeV simulated with 
PYTHIA,  version 6.4, using default parameters \cite{Pythia06}.  PYTHIA (versions 6.2 and 6.3) simulations were found to qualitatively reproduce 
inclusive and identified particle spectra measured by STAR in $p+p$ collisions at $\sqrt{s}=200$ GeV \cite{Abelev2007:PRC75,Adams2006:PRD74}, as well particle yield ratios \cite{Adams2006:PLB637}. 
Figures~\ref{fig:3}, \ref{fig:5}, and \ref{fig:7} present two-particle correlations in $\Delta \eta$ vs. $\Delta \phi$ while figures \ref{fig:4}, \ref{fig:6}, and \ref{fig:8}
 display three-particle correlations 
in $\Delta \phi_{12}$ vs $\Delta \phi_{13}$  relative azimuthal
angles. 
All correlations are calculated for charged particles produced in 
the range $|\eta|<1$. 
Two-particle correlations are studied between "high" and "low" momenta 
particles. For three particle correlations, we use one high 
momentum "trigger" particle, and two lower 
momenta particles in the same range. 
Specific transverse momentum ranges used in the different 
correlations are discussed next.

We illustrate that the strength and shape of the correlations 
depend on the transverse momentum ranges considered. 
Panel (a) of figures 3, 5, and 7, show two-particle correlations obtained for 
a jet tag particle in the range $3 < p_{t} < 20$~GeV/c, and "associates" in the ranges 
$0.2 < p_{t} < 1$~GeV/c, $1 < p_{t} < 2$~GeV/c, 
and $2 < p_{t} < 3$~GeV/c, respectively. 
The top subpanel of each figure show contour plots of the normalized two-particle 
density, $1/N_{2}^{tot}d^2N_{2}/d\Delta\eta d\Delta\phi$, where $N_{2}^{tot}$ is  the integral of the correlation function.
The middle and bottom subpanels display projections $dN_{2}/d\Delta\eta$ and $dN_{2}/d\Delta\phi$ 
obtained for $|\Delta\phi|<0.7$ rad, and $|\Delta\eta|<1.0$, respectively. 
The vertical axes of the projections are labeled $N_{2}$ for brevity in the figures.

The two-particle densities, and their projections, obtained for different $p_{t}$ ranges exhibit 
similar features but with varying strengths and widths.  
The peak centered at $\Delta\eta =0$, $\Delta\phi=0$ corresponds to the 
"triggered" or tagged jet. The narrow ridge-like structure 
centered at  $\Delta\phi=\pi$ which extends over the 
full $\Delta\eta =2$ range covered by the plot arises from 
the fact that in $p+p$ intereactions, 
jets are produced by partons colliding with typically different 
fractions "x" of the proton momenta.  
Consequently, while the jets are nearly back-to-back in 
azimuth (because of momentum conservation), they may have quite 
different rapidities thereby leading to an elongated away-side
structure as seen in panel (a) of figures \ref{fig:3}, \ref{fig:5}, and \ref{fig:7}.

It is also useful to consider the jet correlation structures measured 
on the basis of three-particle azimuthal correlations to shed some
light on features expected in ongoing search for conical emission 
in heavy ion collisions. Panel (a) of figures \ref{fig:4}, \ref{fig:6}, and \ref{fig:8}   display 
three-particle correlations in the form of 3-cumulants plotted as 
a function of two relative azimuthal angles $\Delta \phi_{12}$ and $\Delta 
\phi_{13}$. 
The middle and bottom subpanels show projections of the
3-cumulant along diagonal axes.
The  middle panels projections are carried along the 3-cumulant main diagonal 
$\Delta \phi_{12}+\Delta \phi_{13}$ with a condition 
$|\Delta \phi_{12}-\Delta \phi_{13}|< 0.35$, and shifted so their 
origin  coincide with the away-side $(\Delta \phi_{12}+\Delta
\phi_{13})/2-\pi=0$. 
The bottom panels show projections of the 3-cumulant along 
the second diagonal $\Delta \phi_{12}-\Delta \phi_{13}$ with 
the condition $|(\Delta \phi_{12}+\Delta \phi_{13})/2-\pi|<0.3$ 
chosen to emphasize the shape and strength of the away-side 
peaks.  The 3-cumulant used here was introduced in reference
\cite{Pruneau06}, and is defined as 
\begin{equation}
 C_3 (\Delta \phi _{12} ,\Delta \phi _{13} ) = \rho _3  - 
\rho _2^{(12)} \rho _1 ^{(3)} 
 - \rho _2^{(13)} \rho _1 ^{(2)}   - \rho _2 ^{(23)} \rho _1 ^{(1)} + 
2\rho _1 \rho _1 \rho _1 
\end{equation}
in terms of the measured 3-particle density (number of 
triplets/event) $\rho_3\equiv dN_3/d\Delta \phi _{12} d\Delta 
\phi _{13}$ and combinatorial terms $\rho _2 ^{(ij)} \rho _1 
^{(k)}$
 and $\rho_1\rho_1\rho_1$ calculated based on measured 2- and 1- 
particle densities. 
The 3-cumulant is used, instead of a simpler 3-particle density  $\rho^{(123)}_3= 
d^2N_3/d\Delta\phi_{12}d\Delta\phi_{13}$, because it enables a measurement of three correlated 
particles by eliminating the combinatorial contributions $\rho^{(ij)}_2\rho^{(k)}_1$ where  
$\rho^{(ij)}_2 = dN_2/d\Delta\phi_{ij}$ and $\rho^{(k)}_1 = dN_1/d\phi_{k}$.

The 3-cumulants shown in panel (a) of figures \ref{fig:4}, \ref{fig:6}, \ref{fig:8} are obtained with the same 
pseudo-rapidity, and transverse momentum cuts as those used 
in Figures \ref{fig:3}, \ref{fig:5}, \ref{fig:7}. The high momentum cut defines particle "1", while 
the low momentum cut is used to select particles "2" and "3".  
The 3-cumulants exhibit peak structures 
expected for two jets produced nearly back-to-back in 
azimuth \cite{Pruneau06}. The peak centered at $\Delta \phi_{12}=0, 
\Delta \phi_{13}=0$ corresponds to two "low" momentum particle 
being emitted along side the "trigger" high $p_{t}$ particle. 
The peak centered at $\Delta \phi_{12}=\pi, \Delta \phi_{13}=\pi$ 
corresponds to two "low" momentum particle being emitted nearly 
back-to-back with the "trigger" high $p_{t}$ particle, and thus 
correspond to an away-side jet. 
Peaks at $\Delta \phi_{12}=\pi (0), \Delta \phi_{13}=0 (\pi)$ 
correspond to combinations where one of the low 
momenta particle is along-side the high $p_t$ particle and 
the other back-to-back.  One observes from 
panel (a) of figures \ref{fig:4}, \ref{fig:6}, \ref{fig:8} that the strength of the 3-cumulant decreases progressively for increasing selected associate momenta. 
One additionally finds that peaks are narrower at high momentum 
owing to the fact that high $p_t$ particles tend to be better 
aligned with the direction of the jet they are produced from. 
Also note that the away side peak 
($\Delta \phi_{12}=\pi, \Delta \phi_{13}=\pi$) is wider along the main 
diagonal than the alternate diagonal. This is usually associated 
to the fact that the incoming partons that 
produce the jets are not co-linear in the laboratory reference 
frame, i.e. they carry a small but finite 
transverse momentum: the jets are therefore  not exactly 
back-to-back in azimuth, thereby leading to finite 
elongation of the peak along  the main diagonal 
$\Delta \phi_{12}+ \Delta \phi_{13}$ axis.

\section{The effect of radial flow on two- and three- particle correlations}

We now proceed to study the impact of radial flow on 
two- and three- particle correlations. We  use 
single $p+p$ PYTHIA events radially boosted at fixed velocity in
random directions in the transverse plane to simulate the effect of 
radial flow on particle correlations. The use of a single event presents the 
advantage one does not need to carry background subtraction of combinatorial terms that arise in A+A collisions.
One can picture the phenomenon along two different scenarios:
first consider two incoming partons are 
subject to transverse kicks before they collide to produce jets.  As stated in the introduction, this phenomenon has 
been invoked successfully to interpret Drell Yan data \cite{Gavin:1988tw}, and also contributes to the Cronin 
effect in jets \cite{Accardi:2005fu}. In this scenario, the jet production in the rest frame of the parton pair
 is essentially the same as if the partons were not radially boosted, but the produced hadrons are boosted in 
the lab frame thereby leading to modified correlations as shown in the following. 

In an alternative scenario, one considers  incoming partons interact as if in vacuum, but their collision products are 
subject to a transverse boost due to momentum kicks by the dense medium. This assumes jet hadronization proceeds quickly within the
medium, and sufficient medium flow builds up to radially push hadrons produced by parton fragmentation.
One does not expect all hadrons to be subjected to the same kick but overall to a finite average kick.

We assume that in either case, the particle production, locally 
in the parton-parton rest frame, proceeds similarly as in vacuum, 
so as to preserve hadron correlations. While we do not, in practice, expect the partons to behave quite that way, the above scenarios enable a simple exploration of the effect of radially boosted particles produced a by reasonably well calibrated common source: PYTHIA. We note however that the boosting of an entire PYTHIA event rather than only hadrons from jets results in an artificial enhancement of the near correlations.

In a nucleus-nucleus collision, we would further model the radial 
boost velocity as being proportional to 
some small power of the distance, in the transverse plane, 
between the parton-parton interaction point 
and the nucleon participants center-of-mass position. However, one should also account for quenching of the jets, as well as rescattering and diffusion effects. Such complications are here neglected by considering correlation induced by one radially boosted p+p collision only. Note additionally that in an actual $A+A$ collision, multiple p+p collisions would suffer a radial boost in different directions. As these are to first order uncorrelated, this would lead to a large flat background in the correlations under study. The strength of the correlations above this background shall depend on the number of $p+p$ or "di-jet" sources. Such effects are also neglected.

We study the impact of radial boost in finite increments of velocity 
to understand the modifications imparted to the particle correlations.  
Panels (b), (c), and (d) of figures 4-8 show two- and three- particle correlations obtained with 
radial velocities $\beta_r =$0.2, 0.3 and 0.4 respectively. Rapidity and transverse momentum 
cuts are identical to those used in panel (a) of these figures.

We make the following observations on the basis of the two-particle 
$\Delta\eta$ vs. $\Delta\phi$ correlations. 
First, we note that the near side "jet peak" tends to increase 
in amplitude progressively with larger radial velocity. 
The amplitude increase stems form the kinematical 
focusing produced by the radial boost on the trigger particles and 
its associates: with large $\beta$, more particles are focused on 
the near side peak in the given $p_t$ range of interest.  

Next, we observe the away-side ridge flattens out in azimuth at small 
velocity, progressively separates into two parallel 
structures with increasing velocity, and is refocused on the near 
side at the highest velocities considered here.  
We note the weak away-side "two-bumps" structures seen at low beta 
in projections on the $\Delta\phi$   axis are  qualitatively similar 
to the structures reported by STAR and PHENIX based on 
two-particle correlations \cite{Horner06}. 
The away-side ridge is essentially deflected and defocused away. , 
This leads to an increase of the correlation background for low $\beta_r$ and to progressive build up of a ridge on the near side for larger boost velocities in qualitative agreement with observations by STAR~\cite{Horner06,Putschke06}. 
Radial flow can thus simultaneously produce the disappearance of 
the away-side "jet" peak, the formation of a dip structure, 
and the production of a strong ridge on the 
near side.

We note in closing that given our simulations use one boosted $p+p$ per simulated event, momentum conservation is globally violated. To achieve global momentum conservation, one would require addition of a recoiling background. Such background shall lead to a $cos(\Delta\phi)$ in the correlation functions thereby resulting in an  increase of the away side yield \cite{Borghini06}. The strength of this effect shall depend on event multiplicity, momentum ranges considered, and details of the collision dynamics. Effects of the recoil were therefore neglected in the simple model presented in this paper.     

\section{Discussion}

The model used in this work is obviously simplistic. 
We do not expect parton-parton collisions in a dense-medium 
to proceed exactly as those in vacuum.
We also do not expect all particles produced at a given radius in A+A 
to be subject to equal radial collective velocity. 
The use of radially boosted PYTHIA final 
states is thus not entirely justified. 
The spirit of the model should however be considered seriously as it 
provides a simple and elegant explanation to several
 observed phenomena simultaneously. A quantitative 
approach however requires attention to details: 
one must indeed consider modifications of the cross-
section in the presence of the dense medium medium, gluon radiation, 
longitudinal flow effects, and rescattering or diffusion effects.

Explaining the large energy and particle yield of the jet in terms of 
energy loss is also problematic given the observation that the ridge 
yield is basically independent of the selected trigger particle pt 
range, while the yield of the jet peak itself is independent of the selected pt. 
The radial boost mechanism, on the other hand, does not specifically 
modify the jet itself, but enables the formation of a strong ridge 
correlated in azimuth to the jet. The ridge thus carries high particle 
multiplicity yields, and energy, by construction.

Acknowledgements

The authors acknowledge support by DOE Grant No. DE-FG02-92ER40713.


\section*{References}
\begin{thebibliography}{10}

\bibitem{Ackermann01} K. H. Ackermann, {\em et al.}, (STAR Collaboration),  Phys. Rev. Lett.  86, 
402 (2001). 

\bibitem{Adler03} C. Adler, et al. STAR Collaboration, Phys. Rev. Lett.  90, 082302 (2003).

\bibitem{Horner06} M. Horner, {\em et al.} (STAR Collaboration), 
Proceedings of the Quark Matter 06 Conference, Shanghai, China, 2006, J. of Physics G: Nuclear and Particle Physics, in press.


\bibitem{Putschke06} J.  Putschke, {\em et al.} (STAR Collaboration), Proceedings of QM06; 
arXiv:nucl-ex/0701074,  2007.

\bibitem{Star05} J. Adams {\em et al.} (STAR Collaboration), Phys. Rev. Lett. {\bf 95}, 152301 
(2005); 





\bibitem{Ulery05} J. Ulery, {\em et al.} (STAR Collaboration), Nucl. Phys. A {\bf 774}, 581 (2006).

\bibitem{Phenix06} S. S. Adler {\em et al.} (PHENIX Collaboration), Phys. Rev. Lett. 97, 052301 
(2006).
\bibitem{Stoecker05} H. Stoecker, Nucl. Phys. {\bf A750}, 121 (2005).
\bibitem{Solana05} J. Casalderrey-Solana {\em et al.} J. Phys. Conf. Ser. {\bf 27}, 23 (2005).
\bibitem{Ruppert05} J. Ruppert, B. Muller, Phys. Lett. B {\bf 618}, 123 (2005).

\bibitem{Stocker07} H. Stocker , {\em et al.} arXiv:nucl-th/0703054, 2007.

\bibitem{Heinz06} A. K. Chaudhuri, U. W. Heinz, Phys. Rev. Lett. {\bf 97}, 062301 (2006).
\bibitem{Vitev05} I. Vitev, Phys. Lett. B {\bf 630}, 78 (2005).

\bibitem{Salgado05}  A.D. Polosa and C.A. Salgado, hep-ph/0607295. 

\bibitem{Dremin05}  I.M. Dremin, Nucl. Phys. A {\bf 767}, 233 (2006).

 \bibitem{Majumber05}  V. Koch, A. Majumber, X.-N. Wang, Phys. Rev. Lett. {\bf 96}, 172303 
(2006). 

\bibitem{Salgado05a} N. Armesto, C.A. Salgado, U. A. Wiedemann, Phys. Rev. C {\bf 72}, 064910 
(2005).





\bibitem{Ma06} Y. G. Ma, {\em et al.} arXiv:nucl-th/0610051, arXiv:nucl-th/0608050, arXiv:nucl-
th/0610088. 

\bibitem{Pruneau06} C. Pruneau, Phys. Rev. C 74, 064910 (2006), arXiv:nucl-ex/0608002.


\bibitem{Borghini06} N. Borghini, nucl-th/0612093.

\bibitem{Ajitanand06} Chun Zhang {\em et al.} (PHENIX Collaboration), Quark Matter 2006 - The 19th International Conference on Ultra-Relativistic Nucleus-Nucleus Collisions, Shanghai, China, Nov 2006,  Journal Physics G 34, s671 (2007).

\bibitem{Ulery06} J. Ulery and F. Wang, nucl-ex/0609016.

\bibitem{Pruneau06a} C. Pruneau, {\em et al.} (STAR Collaboration), 
Quark Matter 2006 - The 19th International Conference on Ultra-Relativistic Nucleus-Nucleus Collisions, Shanghai, China, Nov 2006,  Journal Physics G 34, s667 (2007).

\bibitem{CeresQM06} Stefan Kniege and Mateusz Ploskon, {\em et al.} (CERES  Collaboration), Quark Matter 2006 - The 19th International Conference on Ultra-Relativistic Nucleus-Nucleus Collisions, Shanghai, China, Nov 2006,  Journal Physics G 34, s697 (2007).

\bibitem{Magestro04} D. Magestro, {\em et al.} (STAR Collaboration), Proceedings of Hard Probe 
Conference, 2004.



\bibitem{Armesto04}  N. Armesto, {\em et al.}, Phys. Rev. Lett. {\bf 93}, 242301 (2004), hep-
ph/0405301.
\bibitem{MajumderMuller06} A. Majumder, B. Muller, S. A. Bass (2006), hep-ph/0611135.
\bibitem{Romatschke07} P. Romatschke, Phys. Rev. C{\bf 75}, 014901 (2007), hep-ph/0607327.

\bibitem{Voloshin05} S.A. Voloshin, Phys. Lett. B {\bf 632}, 490 (2006) [arXiv:nucl-th/0312065].

\bibitem{Hwa05} C. B. Chiu, R. C. Hwa, Phys. Rev. C{\bf 72}, 034903 (2005), nucl-th/0505014.

\bibitem{Magestro06} D. Magestro, {\em et al.} (STAR Collaboration), Nucl.Phys. A{\bf 774} (2006) 573-576.
\bibitem{Magestro06a} 
J. Adams, {\em et al. }, Phys. Rev. Lett. {\bf 97} (2006) 162301.
\bibitem{EllipticalFlow} J. Adams, {\em et al. }, (STAR Collaboration) Nucl. Phys. A {\bf 757} (2005) 102. 
K. Adcox, {\em et al. }, (PHENIX Collaboration) Nucl. Phys. A {\bf 757} (2005) 184. 


\bibitem{MolnarBarannikova} L. Molnar and O. Barannikova, and F. Wang, Romanian Reports in Physics, {\bf 58}, 19, 2006.
\bibitem{RetiereLisa} F. Retiere and M. Lisa, Phys. Rev. C {\bf 70} (2004). 044907
\bibitem{Voloshin06} S. A. Voloshin, Nucl. Phys. A{\bf 749}, (2005) 287.

\bibitem{WhitePapers}
  I.~Arsene {\it et al.}  [BRAHMS Collaboration],
   Nucl.\ Phys.\ A {\bf 757}, 1 (2005);  K.~Adcox {\it et al.}  [PHENIX Collaboration],
   {\it ibid}.~184;   B.~B.~Back {\it et al.} [PHOBOS Collaboration],
  {\it ibid}.~28;
  J.~Adams {\it et al.}  [STAR Collaboration],
  {\it ibid}. ~102.

\bibitem{Huovinen:2006jp}
  P.~Huovinen and P.~V.~Ruuskanen,
  Ann.\ Rev.\ Nucl.\ Part.\ Sci.\  {\bf 56}, 163 (2006)
  [arXiv:nucl-th/0605008].


\bibitem{Zhang:2005ni}
  B.~Zhang, L.~W.~Chen and C.~M.~Ko,
  Phys.\ Rev.\  C {\bf 72}, 024906 (2005)
  [arXiv:nucl-th/0502056].

\bibitem{adare2007} A. Adare, {\it et al.}, Phys. Rev. Lett. {\bf 98}, 172301 (2007).


\bibitem{Gavin:1988tw}
  S.~Gavin and M.~Gyulassy,  Phys.\ Lett.\  B {\bf 214}, 241 (1988).

\bibitem{Guo:1999wy}
  X.~f.~Guo, J.~w.~Qiu and X.~f.~Zhang, Phys.\ Rev.\ Lett.\  {\bf 84}, 5049 (2000)
  [arXiv:hep-ph/9911476].


\bibitem{Alde:1991sw}
P.~Bordalo {\it et al.}  [NA10 Collaboration], Phys.\ Lett.\  B {\bf 193}, 373 (1987); D.~M.~Alde {\it et al.}, Phys.\ Rev.\ Lett.\  {\bf 66}, 2285 (1991); Phys.\ Rev.\ Lett.\  {\bf 66}, 133 (1991).

\bibitem{Accardi:2005fu}
  A.~Accardi, Eur.\ Phys.\ J.\  C {\bf 43}, 121 (2005)
  [arXiv:nucl-th/0502033]; arXiv:hep-ph/0212148.

\bibitem{Fries:2006pv}
 R.~J.~Fries, J.~I.~Kapusta and Y.~Li, arXiv:nucl-th/0604054.

\bibitem{RenkRupper:2005}
  T. Renk and J. Ruppert, `
  Phys. Rev. C{\bf 72}, 044901 (2005).


\bibitem{Pythia06} T. Sjšstrand, S. Mrenna and P. Skands, J. High Energy Phys. JHEP05 (2006) 026.

\bibitem{Abelev2007:PRC75} B.I.Abelev (STAR Collaboration), Phys. Rev. C{\bf 75} (2007), 064901.
\bibitem{Adams2006:PRD74}  J.Adams (STAR Collaboration), Phys. Rev. D{\bf 74} (2006), 032006.
\bibitem{Adams2006:PLB637} J.Adams (STAR Collaboration), Phys. Lett. B{\bf 637} (2006), 161.

\end{thebibliography}
\end{document}